\documentclass{IEEEoj}
\usepackage{graphicx}
\usepackage{svg}
\usepackage{comment}
\usepackage{bbding}
\usepackage{multirow}
\usepackage{longtable}
\usepackage{array}
\usepackage{cite}
\usepackage{amsmath}
\usepackage{amssymb}
\usepackage{hyperref}
\usepackage{indentfirst}
\usepackage{verbatim}
\usepackage{makecell}
\usepackage{soul}
\usepackage{tabularx}
\usepackage{booktabs}
\usepackage{subcaption}
\usepackage{lettrine}
\usepackage{algorithm}
\usepackage{algorithmic}
\usepackage{colortbl} 
\usepackage{subcaption}
\newtheorem{definition}{Definition}

\usepackage[shortcuts,acronym, nonumberlist]{glossaries}

\newcommand{\Rev}[1]{{\color{black}{#1}}}
\def\BibTeX{{\rm B\kern-.05em{\sc i\kern-.025em b}\kern-.08em
    T\kern-.1667em\lower.7ex\hbox{E}\kern-.125emX}}
\AtBeginDocument{\definecolor{ojcolor}{cmyk}{0.93,0.59,0.15,0.02}}
\def\OJlogo{\vspace{-4pt}\hskip-4pt\includegraphics[height=18pt]{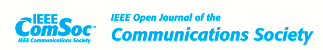}}

\makeglossaries 
\newacronym{cnn}{CNN}{Convolutional Neural Network}
\newacronym{dnn}{DNN}{Deep Neural Network}
\newacronym{patchtst}{PatchTST}{Patch Timeseries Transformer}
\newacronym{dl}{DL}{Deep Learning}
\newacronym{iq}{IQ}{In-phase and Quadrature}
\newacronym{ssl}{SSL}{Semi-Supervised Learning}
\newacronym{dae}{DAE}{Deep Autoencoders}
\newacronym{snr}{SNR}{Signal-To-Noise Ratio}
\newacronym{stft}{STFT}{Short-Time Fourier Transform}
\newacronym{mtf}{MTF}{Markov Transition Field}
\newacronym{wtr}{WTR}{Wireless Technology Recognition}
\newacronym{ml}{ML}{Machine Learning}
\newacronym{rnn}{RNN}{Recurrent Neural Networks}
\newacronym{lstm}{LSTM}{Long Short-Term Memory}
\newacronym{los}{LOS}{Line-Of-Sight}
\newacronym{nlos}{NLOS}{Non-Line-Of-Sight}
\newacronym{cir}{CIR}{Channel Impulse Response}
\newacronym{mse}{MSE}{Mean Squared Error}
\newacronym{uwb}{UWB}{Ultra-Wideband}
\newacronym{mae}{MAE}{Mean Absolute Error}

\begin{document}
\setstcolor{red}
\receiveddate{XX Month, XXXX}
\reviseddate{XX Month, XXXX}
\accepteddate{XX Month, XXXX}
\publisheddate{XX Month, XXXX}
\currentdate{11 January, 2024}
\doiinfo{OJCOMS.2024.011100}

\title{A Unified Foundation Model for Wireless Technology Recognition and Localization}

\author{Mohammad~Cheraghinia\IEEEauthorrefmark{1}, Eli~De~Poorter \IEEEauthorrefmark{1},         Jaron~Fontaine \IEEEauthorrefmark{1}, 
Merouane~Debbah \IEEEauthorrefmark{2},
Adnan~Shahid \IEEEauthorrefmark{1}
}
\affil{IDLab, Department of Information Technology, Ghent University - imec, 9052 Ghent, Belgium.}
\affil{Center for 6G Technology, Khalifa University of
Science and Technology, P O Box 127788, Abu Dhabi, United Arab Emirates.}
\corresp{CORRESPONDING AUTHOR: Mohammad~Cheraghinia (e-mail: mohammad.cheraghinia@ugent.be).}
\authornote{This work was partly funded by the SNS JU European Union’s Horizon Europe research and innovation program under Grant Agreement No 101139194 (6GXCEL), the Horizon Europe program under the MCSA Staff Exchanges Grant Agreement 101086218 (EVOLVE), and the Fund for Scientific Research Flanders (FWO-Vlaanderen) under SB-PhD Fellowship with grant number 1S52025N and FWO research project PESSO with grant number G018522N.}
\markboth{Preparation of Papers for IEEE OPEN JOURNALS}{Author \textit{et al.}}

\begin{abstract}
\Rev{\ac{wtr} and localization are essential in modern communication systems, enabling efficient spectrum management, seamless coexistence of diverse technologies, and accurate positioning in dynamic environments. In real-world conditions, solutions must handle signals from various resources with different sampling rates, capturing devices, frequency bands, and propagation conditions. Traditional methods, such as energy detection and conventional \ac{dl} models like \glspl{cnn}, often lack the robustness to generalize across unseen technologies, environments, or tasks. In this work, we introduce a Transformer-based foundation model for both WTR and localization, pre-trained in a self-supervised manner on large-scale, unlabeled datasets of \ac{iq} and \ac{cir} timeseries. The model leverages input patching for computational efficiency and employs a two-stage pipeline: self-supervised pre-training to learn general-purpose representations, followed by lightweight fine-tuning for task-specific adaptation. This enables the model to generalize to new wireless technologies and unseen environments using minimal labeled samples. Evaluations across short-range and long-range datasets show superior accuracy in WTR (up to 99.99\%), \ac{los} detection (up to 100\%), and ranging error correction (reducing \ac{mae} by up to 50\%), all while maintaining low computational complexity. These results underscore the potential of a reusable wireless foundation model for multi-task applications with minimal retraining.}
\end{abstract}


\maketitle

\section{INTRODUCTION }
\Rev{The evolution of wireless communication systems has shown an era of unique connectivity, including applications from autonomous vehicles \cite{8809282} and smart homes \cite{7757102} to localization ecosystems \cite{10738385}. Wireless systems must handle the coexistence of diverse technologies, including long-range technologies (e.g., Sigfox, LoRa, IEEE 802.11ah, IEEE 802.15.4) for energy-efficient coverage and short-range technologies (e.g., LTE, WiFi, 5G) for high-data-rate links. Effective \ac{wtr} specifies signals based on features like bandwidth, modulation, and temporal patterns, which is important for spectrum management \cite{8325299}, interference mitigation \cite{7289294}, and secure spectrum sharing \cite{8281460}. On the other hand, precise localization in multipath environments, such as distinguishing between \ac{los} and \ac{nlos} paths in \ac{uwb} systems and correcting ranging errors, is essential for reliable positioning in dynamic settings, such as warehouses or offices.

However, real-world wireless solutions face challenges due to data heterogeneity: signals vary in sampling rates, capture devices, frequency bands, and environmental conditions, leading to poor generalization in traditional models \cite{zhu2025wirelesslargeaimodel}. Task-specific approaches often require large labeled datasets for each scenario separately, which fail to generalize to unseen technologies, devices, or environments without extensive retraining. This inefficiency is worsened by the growing complexity of wireless applications, where new standards emerge rapidly and environments change unpredictably \cite{pmlr-v235-goswami24a}.

To overcome these limitations, foundation models are pre-trained on vast amounts of unlabeled data to learn general-purpose features, establishing a transformative paradigm. Inspired by models in natural language processing \cite{NIPS2017_3f5ee243} and computer vision \cite{10834497}, foundation models enable learning across diverse tasks, reducing dependence on task-specific labeled data and enabling generalization to novel scenarios. In wireless communications, a foundation model is required to do various tasks such as \ac{wtr} and localization, capturing patterns in raw \ac{iq} and \ac{cir} timeseries. Why is this essential? First, it enables data efficiency: by pre-training on diverse unlabeled signals, the model learns robust features that generalize to unseen classes or environments, thus overcoming the challenge of limited data availability in wireless datasets. Second, it enhances scalability: as the number of wireless technologies increases (e.g., toward 6G and beyond), a reusable foundation can adapt to new technologies without rebuilding models from scratch, aligning with ongoing research in multi-task learning. Third, it supports computational sustainability: through techniques like patching, it handles long sequences efficiently, avoiding the complexity of vanilla Transformer architectures while preserving dependencies. }

\Rev{In this work, we propose a Transformer-based foundation model pre-trained in a self-supervised manner on \ac{iq} and \ac{cir} datasets. Our direct timeseries solution bypasses the computational overhead associated with spectrogram generation, using the critical temporal and phase features from raw \ac{iq} and \ac{cir} data. Unlike spectrograms, which require an additional transformation step, this raw data is readily available from many commercial radio front-ends (e.g., UWB transceivers like the Qorvo DW3000 series \cite{QorvoDWM3000} or various software-defined radios \cite{5462981}), making our solution directly applicable to existing hardware. Using patching for reduced complexity and a two-stage pipeline (self-supervised pre-training followed by lightweight fine-tuning), the model generalizes across short-range and long-range technology recognition tasks, as well as UWB localization tasks ((N)LOS classification and localization error correction). To assess our model under challenging conditions, we employ different datasets to include sampling rate and capturing device diversity. Furthermore, we evaluate its generalization capabilities by testing it in different environments for localization tasks and on unseen classes for the WTR tasks. Experimental results demonstrate its effectiveness: achieving up to 99.99\% accuracy in multi-class WTR, 100\% in LOS/NLOS detection, and up to 50\% reduction in ranging \ac{mae}.

The main contributions of the paper are summarized as follows:
\begin{itemize}

\item We propose a foundation model that directly uses raw timeseries data (\ac{iq} and \ac{cir}), eliminating the overhead of data pre-processing. To our knowledge, this is the first wireless foundation model capable of processing heterogeneous input types like IQ samples and CIR (each with distinct data features) within a single, unified architecture.

\item We propose a patching strategy designed for our Transformer architecture. This technique addresses the challenge of processing sequences by segmenting the input data into smaller patches, enabling the model to enhance its computational efficiency and accuracy.

\item We introduce a foundation model that adapts across \ac{wtr} and localization tasks by handling heterogeneous data from \textbf{multiple technologies, sampling rates, and environments}. Through an evaluation of pre-training strategies, we demonstrate that a self-supervised approach achieves competitive or superior performance to supervised methods, validating the model's effectiveness as a true wireless timeseries foundation model.

\item We conduct an evaluation across a diverse range of downstream tasks, encompassing both \textbf{classification and regression}. Using short-range, long-range, and UWB datasets, we assess the model's generalization capabilities, specifically its performance in \textbf{unseen environments and wireless technologies} not present during pre-training. Our model is benchmarked against baselines across multiple metrics, including predictive accuracy and model complexity.
    
\end{itemize}

The remainder of the paper is organized as follows: Section \ref{sec:related} reviews related works and explains the wireless foundation models and related studies. Section \ref{sec:system} introduces the system model and description of pre-training and fine-tuning strategies. Section \ref{sec:method} describes the methodology, including the architecture of the proposed solution and baseline models. Section \ref{sec:results} presents results, discussion, and computational analysis. Finally, Section \ref{sec:conclusion} concludes the paper and provides directions for future work.}

\section{Related Works} \label{sec:related}

\Rev{
This section is an overview of the literature relevant to our work, including three key domains. We first review the application of AI in \ac{wtr} and \ac{uwb} localization, exploring both established and learning-based methods in comparison to our downstream tasks. We then delve into the evolving field of foundation models for wireless applications, providing the direct context for our proposed approach.

\subsection{AI in WTR}
Traditional methods, such as energy detection, are simple but struggle in noisy and complex environments, while more robust techniques, like cyclostationary detection, are computationally expensive \cite{s17092081, 5072368}. \glspl{cnn} represent a major improvement by automatically learning features from raw \ac{iq} data. Studies show \glspl{cnn} can achieve high accuracy 99\% on low-power devices \cite{9348566}. To address the large data requirements of \glspl{cnn}, \ac{ssl} has been used to achieve over 70\% accuracy with only 10\% of the labeled data \cite{8935690}. However, a key weakness of \glspl{cnn} is their poor ability to generalize to new, unseen wireless technologies without extensive retraining, as they rely on convolutional kernels to capture local patterns from the training data\cite{5629572, 8824856, FONTAINE2019101881}.

Transformers are well-suited for sequence data like \ac{iq} samples due to their ability to model timseries dependencies. Several studies \cite{9674558, 10093837, electronics13173408, 9779340} have applied Transformers to \ac{iq} samples using various techniques like hybrid architectures, \ac{ssl}, and different data windowing methods. While promising, these approaches often focus on modulation classification in simulated environments and do not consider data with different sampling rates and capturing devices. Vision-based models convert wireless signals into images (e.g., spectrograms) and then apply computer vision techniques. These methods have shown high accuracy \cite{9826820, 10195842} but can be computationally heavy, requiring pre-processing time and very large models with millions of parameters \cite{aboulfotouh20256gwavesfmfoundationmodel}.

Recent studies have explored self-supervised and meta-learning solutions to remove the need for extensive labeled datasets \cite{10049409, 9930826, 10901963}. Some approaches convert timeseries signals into image-based representations, like time-frequency diagrams \cite{10049409}, or utilize multi-modal inputs (e.g., IQ, Amplitude/Phase, and Power Spectral Density) to tackle few-shot or open-set challenges \cite{10949587}. While effective, these methods can introduce computational overhead and associated latency from the pre-processing transformations \cite{10857965}, unlike our approach that uses the direct \ac{cir} or \ac{iq} output of off-the-shelf devices. Other studies apply various contrastive and self-supervised learning frameworks directly to raw timeseries data \cite{9930826, 10901963, 9652914, 9780925}, but they typically focus on the singular task of modulation classification, often with an emphasis on improving robustness against noise \cite{10857965}, channel impairments, developing novel data augmentations \cite{9930826}, or handling out-of-distribution signals \cite{10901963}. Our work differs by proposing a versatile, Transformer-based foundation model pre-trained on raw \ac{iq} and \ac{cir} timeseries. This approach is designed to handle multiple downstream tasks with heterogeneous data and not only enables multi-task capabilities but also avoids the computational limitations and associated delay of image conversion, extending beyond \ac{wtr} to include localization and \ac{los} detection. While this direct timeseries processing may ignore some explicit frequency information present in spectrograms, it preserves the critical phase and temporal details necessary for localization tasks, offering a more efficient and broadly applicable solution.}

\Rev{\subsection{AI in UWB Localization}
Several studies use \ac{ml} to enhance \ac{uwb} positioning accuracy by correcting ranging errors and \ac{los} detection, mainly using the \ac{cir}. These approaches can be grouped into two main categories.

Many studies focus on directly predicting and correcting the ranging error. By extracting features such as signal strength from the CIR, models like Support Vector Machines (SVM) \cite{6192275}, Gaussian Processes, and autoencoders \cite{9151945} have been used for error correction. On the other hand, many studies focus solely on \ac{los} detection. Raw \ac{cir}-based approaches generally show superior performance for (N)LOS detection, with one study reporting a 27.9\% increase in accuracy by using a CNN \cite{9142254}. Similarly, other researchers have combined different neural network architectures to improve performance. \cite{LIU2022101695} applied a Gate Recurrent Unit (GRU) to extract spatial features from the raw \ac{cir} data before feeding them to a CNN for classification. Another hybrid model was proposed \cite{9108193}, which used a CNN to extract non-temporal features that were then fed into an \ac{lstm} network for the final NLOS classification, achieving an accuracy of 82.14\%. 

Other papers first classify the \ac{los} detection and then apply a targeted error correction. Techniques like ensemble tree classifiers, fuzzy logic, and \glspl{cnn} have been employed for the classification step. One study achieved 95.65\% accuracy in (N)LOS detection and reduced the final error to a root mean square error of just 0.4790 m \cite{rayavarapu2022nlos}. Another approach combined ML-based error correction with optimal anchor selection to reduce positioning error by 75\% \cite{9662553}. A critical limitation noted across all these papers is that their methods were not evaluated in new, unseen environments. This makes it difficult to determine how well these solutions would generalize to real-world deployment scenarios. To address this, the study by \cite{10195942} proposes an automatic transfer learning framework that adapts a pre-trained neural network to new environments using a small number of data samples, significantly improving accuracy with minimal new data collection.}

\begin{table*}[h]
\centering
\begin{tabular}{|m{1.9cm}|m{1.4cm}|m{2.9cm}|m{7cm}|m{1.6cm}|}
\hline
\Rev{Model} & \Rev{Input Data} & \Rev{Preprocessing Required} & \Rev{Downstream Tasks} & \Rev{Tasks Type}\\
\hline \hline
\Rev{LWLM~\cite{pan2025largewirelesslocalizationmodel}} & \Rev{CSI} & \Rev{Yes (spatial-frequency)} & \Rev{ToA/AoA estimation, single-base station localization} & \Rev{Regression} \\ \hline
\Rev{IQFM~\cite{mashaal2025iqfmwirelessfoundationalmodel}} & \Rev{IQ} & \Rev{Minimal (augmentations)} & \Rev{modulation classification, AoA, beam prediction, RF fingerprinting} & \Rev{Classification}  \\ \hline
\Rev{WiFo~\cite{liu2025wifo}} & \Rev{CSI} & \Rev{Yes (3D patching)} & \Rev{Channel prediction (time/freq.)} & \Rev{Regression}  \\ \hline
\Rev{SpectrumFM~\cite{zhou2025spectrumfmfoundationmodelintelligent}} & \Rev{IQ} & \Rev{Yes (CNN/MHSA)} & \Rev{modulation classification, wireless technology classification, spectrum sensing, anomaly detection} & \Rev{Classification}  \\
\hline \hline
\Rev{Ours} & \Rev{IQ and CIR} & \Rev{None} & \Rev{wireless technology recognition (2 tasks), LOS detection, ranging error correction} & \Rev{Classification \& Regression}  \\
\hline
\end{tabular}
\caption{\Rev{Comparison of Wireless Foundation Models}}
\label{tab:comparison}
\end{table*}

\Rev{\subsection{Foundation models}

Foundation models have gained attention for learning generalizable representations from diverse datasets, enabling adaptation to multiple downstream tasks. Recent works include Large Wireless Localization Model (LWLM)~\cite{pan2025largewirelesslocalizationmodel}, which pre-trains on Channel State Information (CSI) data using masked modeling and contrastive objectives for tasks like Time-of-Arrival (ToA) estimation, Angle-of-Arrival (AoA) estimation, single-Base Station (BS) localization, and multi-BS localization. IQFM~\cite{mashaal2025iqfmwirelessfoundationalmodel} operates on \ac{iq} streams with contrastive \ac{ssl} and task-aware augmentations, supporting modulation classification, AoA prediction, beam prediction, and Radio Frequency (RF) fingerprinting. WiFo~\cite{liu2025wifo} employs masked autoencoders on CSI for unified time- and frequency-domain channel prediction. SpectrumFM~\cite{zhou2025spectrumfmfoundationmodelintelligent} uses IQ data with masked reconstruction and prediction tasks for modulation classification, wireless technology classification, spectrum sensing, and anomaly detection. 

Unlike these, our model directly processes raw \ac{iq} and \ac{cir} timeseries without preprocessing (e.g., spectrogram or CSI conversion), avoiding pre-processing overhead and preserving critical phase and temporal details essential for localization tasks. IQ samples represent complex-valued raw signal amplitudes over time, capturing modulation and phase information, while CIR timeseries reflect multipath channel delays and amplitudes in the time domain; including these heterogeneous data types in a single Transformer-based model enables unified representation learning across diverse signal types. In contrast, previous models rely on uniform data types (e.g., CSI in LWLM and WiFo, or IQ in IQFM and SpectrumFM) for their respective downstream tasks, limiting adaptability in multi-modal wireless scenarios. This enables better generalization to unseen technologies and environments with minimal labeled data, outperforming baselines in multi-task efficiency.

To highlight differences, Table~\ref{tab:comparison} compares key aspects, showing our compared with the other wireless foundation model studies.}

\begin{figure*}[t]
\centering
\includegraphics[width=0.99\textwidth]{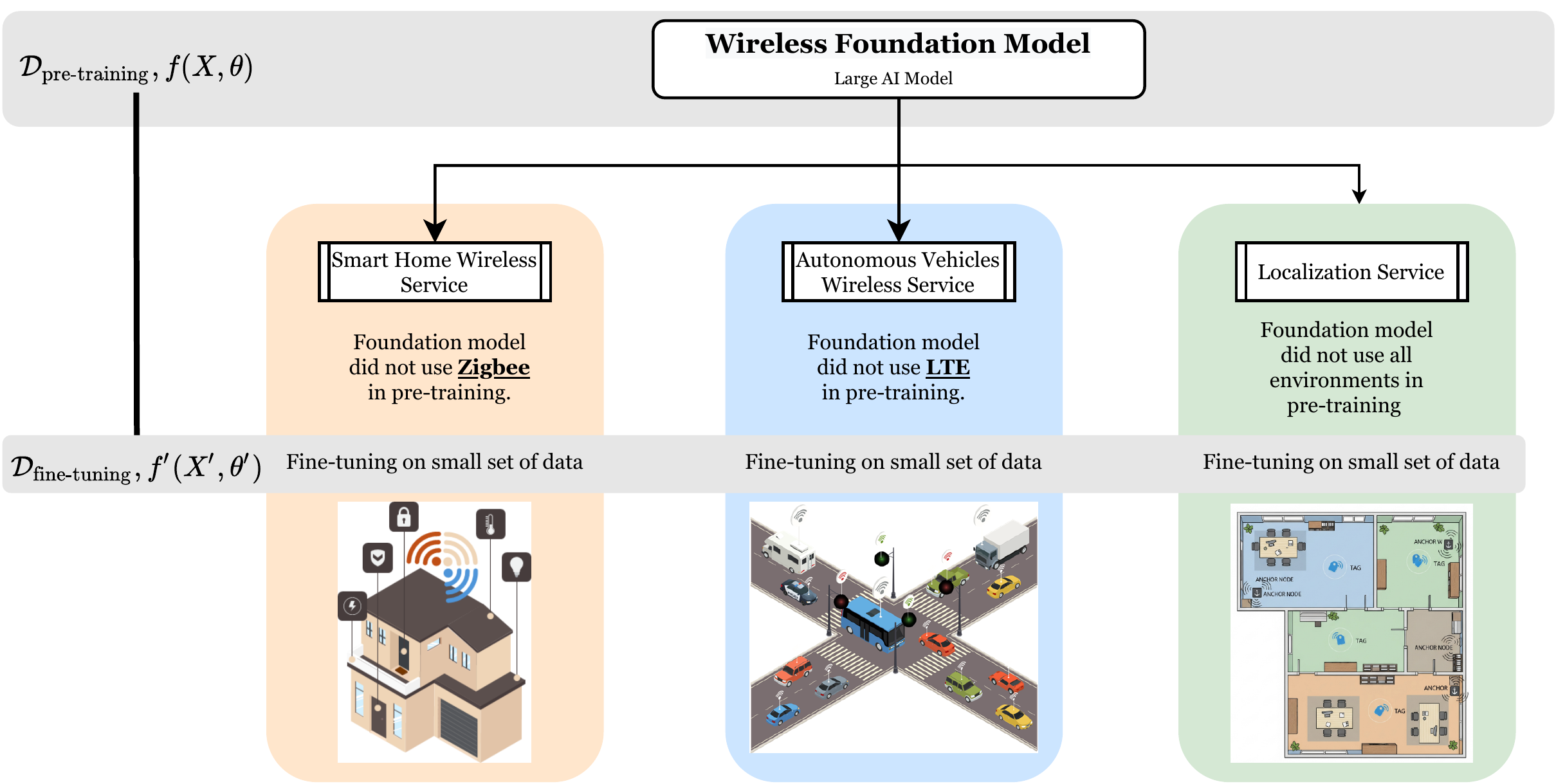}
\caption{\Rev{System model for wireless technology recognition and localization. Fine-tuning necessity on unseen classes in different wireless applications and generalization necessity in diverse environments for localization applications.}}
\label{System:model}
\end{figure*}

\section{System Description}\label{sec:system}
In our proposed system model, a central large model is first pre-trained on a dataset containing long-range (e.g., Sigfox, LoRA, 802.11ah, Zigbee), short-range (e.g., LTE, WiFi, 5G) radio technologies\Rev{, and UWB \ac{los} detection and error correction in different environments (e.g., Office, Hallway, lab, Industrial lab (IIOT))}. \Rev{To focus our pre-training on applicable patterns in each downstream task, we excluded the following technologies and environments in pre-training:}
\begin{enumerate}
    \item \textbf{Zigbee}, which enables low-power and short-range communication suitable for dense sensor deployments in smart home environments~\cite{7757102}.
    \item \textbf{LTE}, which offers wide-area, high-reliability connectivity required for autonomous vehicle operations and vehicular communication~\cite{8809282}. 
    \item \Rev{ \textbf{Localization environments} have different device settings and multi-path patterns. We consider including office and IIOT in the pre-training.}
\end{enumerate}

As depicted in Fig.~\ref{System:model}, the \Rev{foundation model} is then fine-tuned to include these unseen technologies \Rev{and environments} using a small set of labeled data samples related to each application. This fine-tuning process ensures that the model can extend its functionality to previously unseen classes using limited data, improving its generalization across different use cases.

\vspace{1mm}
We define the pre-trained and fine-tuned models as follows (modified from~\cite{HAN2021225}):

\begin{definition}[Pre-trained Model] A \emph{pre-trained model} is a \ac{ml} model trained on a large-scale dataset to learn general representations and serve as a foundation for downstream tasks. \end{definition}

\begin{definition}[Fine-tuned Model] A \emph{fine-tuned model} starts from a pre-trained model and is further trained on a specific labeled dataset to adapt its parameters for a particular task or domain. \end{definition}

\subsection{Problem Statement}
In this work, we explore two distinct pre-training strategies (a) supervised and (b) self-supervised learning, to investigate their impact on downstream performance. This subsection is organized into two parts: the first covers the pre-training phase, divided into separate parts for supervised and self-supervised methods; and the second addresses the fine-tuning phase.

\subsubsection{Pre-training}
We examine two strategies for pre-training the model—supervised and self-supervised learning:

\paragraph{Supervised Pre-training}
We define the \textbf{supervised pre-training dataset} as:

\begin{multline} \label{eq:train_dataset}
\mathcal{D}_{\text{sup}} = \left\{ \left( X_{i_k}, y_k \right) 
\,\middle|\, \right. \\
\quad \left. 
\begin{aligned}
    y_k &\in \mathcal{Y}_{\text{sup}} = \{1, \dots, K\}, \\
    i_k &\in \{1, \dots, M_k\}
\end{aligned}
\right\},
\end{multline}

\noindent where \( X_{i_k} \) represents the data samples belonging to the \( k \)-th class, and this class contains a total of \( M_k \) data samples and \( y_{k} \) denotes their respective labels. Here, \( \mathcal{Y}_{\text{sup}} = \{1, 2, \dots, K\} \) is the set of classes present during pre-training and $K$ is the total number of the classes.

The model \( f(X; \theta) \) is trained to minimize the classification loss. During this stage, the objective is to learn parameters \( \theta \) that capture features and patterns within the classes available in the pre-training set and can be expressed as:

\begin{equation} \label{eq:train_model}
\theta^{\star} = \arg\min_{\theta} \mathcal{L}_{\text{sup}}(\mathcal{D}_{\text{sup}}; \theta).
\end{equation}

\( \mathcal{L}_{\text{sup}} \) is the training loss function can be written as \cite{pmlr-v202-mao23b}:

\begin{multline} \label{eq:train_loss}
\mathcal{L}_{sup}(\theta) = 
    -\sum_{(X_{i_k}, y_{k}) \in \mathcal{D}_{sup}} 
    \sum_{k=1}^{K} \mathbb{1}(y_{k} = k) \\
    \times \log\left(f_k(X_{ i_k}; \theta)\right),
\end{multline}

\noindent where \( f_k(X_{i_k}; \theta) \) denotes the model’s predicted probability for class \( k \) given input \( X_{i_k} \), and \( \mathbb{1}(y_{k} = k) \) is an indicator function that is 1 if \( y_{k} = k \) and 0 otherwise.

\paragraph{Self-supervised Pre-training}

The self-supervised component uses self-supervised reconstruction with patch-level masking, where a random subset of timeseries patches is masked during training. This encourages the model to reconstruct missing segments from data and learn robust temporal features.

The self-supervised dataset is defined as:

\begin{equation}
\mathcal{D}_{\text{unsup}} = \left\{ X_n \right\}_{n=1}^{N},
\end{equation}

\noindent where each \( X_n \in \mathbb{R}^{C \times T} \) is a multichannel timeseries signal. A masking function \( \mathcal{M}(\cdot) \) randomly removes a fixed percentage of the non-overlapping patches in \( X_n \), producing a corrupted input:

\begin{equation}
\widetilde{X}_n = \mathcal{M}(X_n).
\end{equation}

For example, with 50\% masking, half of the patches are zeroed out or replaced with a learnable token.

A reconstruction model \( f_{\text{rec}}(\cdot; \theta) \) is then trained to recover the original signal from the masked version:

\begin{equation}
\theta^{\star}_{\text{unsup}} = \arg\min_{\theta} \mathcal{L}_{\text{unsup}}(\mathcal{D}_{\text{unsup}}; \theta),
\end{equation}

\begin{equation}
\mathcal{L}_{\text{unsup}}(\theta) = \sum_{n=1}^{N} \| f_{\text{rec}}(\widetilde{X}_n; \theta) - X_n \|^2,
\end{equation}

\noindent where \( \| \cdot \|^2 \) denotes the \ac{mse} between the reconstructed and original signals. This encourages the model to learn informative and generalizable representations from raw timeseries data without requiring labels.

\Rev{
\subsubsection{Fine-Tuning}

This phase uses a smaller, task-specific labeled dataset, denoted as $\mathcal{D}_{\text{fine-tuning}}$, which is significantly smaller than the pre-training dataset ($\lvert\mathcal{D}_{\text{fine-tuning}}\rvert \ll \lvert\mathcal{D}_{\text{pre-training}}\rvert$). A principle of this stage is the separation of data to prevent leakage and ensure an unbiased evaluation of the model's generalization:
\[
\mathcal{D}_{\text{pre-training}} \cap \mathcal{D}_{\text{fine-tuning}} = \emptyset.
\]

During fine-tuning, the pre-trained model $f(X; \theta)$ is transformed into a fine-tuned model $f'(X'; \theta')$. To preserve general-purpose representations learned during pre-training, we employ a layer-freezing strategy. The initial layers of the encoder are kept frozen, retaining the foundational knowledge from the pre-trained model. Only the final encoder layer and a newly added task-specific head are unfrozen and trained on the new data.

The overall objective is to find the optimal parameters $\theta'^{\star}$ that minimize the task-specific loss on the fine-tuning dataset:
\begin{equation}
{\theta'}^{\star} = \arg\min_{\theta'} \mathcal{L}_{\text{fine-tuning}}(\mathcal{D}_{\text{fine-tuning}}; \theta').
\end{equation}

We consider two distinct fine-tuning tasks: classification and regression.

\paragraph{Classification Fine-Tuning}
For the classification task, the goal is to categorize wireless technologies. The fine-tuning dataset is defined as:
\begin{multline} \label{eq:cls_tuning_dataset}
\mathcal{D}_{\text{cls-fine-tuning}} = \left\{ \left( X'_{i_k}, y'_k \right)
\,\middle|\, \right. \\
\quad \left.
\begin{aligned}
    y'_k &\in \mathcal{Y}_{\text{fine-tuning}} = \{1, \dots, K'\}, \\
    i_k &\in \{1, \dots, M_k'\}
\end{aligned}
\right\},
\end{multline}
where $X'_{i_k}$ are samples from the $k$-th class, $y'_k$ are their labels, $M_k'$ is the number of samples in class $k$, and $K'$ is the total number of classes. The set of classes extends the pre-training classes $K$ with $U$ new, unseen classes, such that $K' = K + U$ and $\mathcal{Y}_{\text{fine-tuning}} = \mathcal{Y}_{\text{pre-training}} \cup \mathcal{Y}_{\text{unseen}}$. The fine-tuning dataset includes a small number of labeled samples from previously seen classes to support continual recognition.

The model's classification head is a linear layer with a softmax activation. The optimization uses the cross-entropy loss:
\begin{multline} \label{eq:cls_tuning_loss}
\mathcal{L}_{\text{cls-fine-tuning}}(\theta') = -\sum_{(X'_{i_k}, y'_{k}) \in \mathcal{D}_{\text{cls}}} \sum_{k=1}^{K'} \mathbb{1}(y'_{k} = k) \\
\times \log(f'_k(X'_{i_k}; \theta')),
\end{multline}
where $f'_k(X'_{i_k}; \theta')$ is the model's predicted probability for class $k$.

\paragraph{Regression Fine-Tuning}
For the regression task, the objective is to predict the localization error. The fine-tuning dataset consists of input samples and their target vectors:
\begin{equation} \label{eq:reg_tuning_dataset}
\mathcal{D}_{\text{reg-fine-tuning}} = \left\{ \left( X'_{i}, \mathbf{z}'_i \right) \right\}_{i=1}^{N'},
\end{equation}
where $X'_{i}$ is the $i$-th input sample, $\mathbf{z}'_i \in \mathbb{R}^d$ is its label (localization error), and $N'$ is the total number of samples.

For this task, the pre-trained model's head is replaced with a new regression head, a linear layer with one output neuron. The optimization objective is to minimize the \ac{mse} between the predicted and true coordinates:
\begin{equation} \label{eq:reg_tuning_loss}
\mathcal{L}_{\text{reg-fine-tuning}}(\theta') = \frac{1}{N'} \sum_{i=1}^{N'} \| f'(X'_{i}; \theta') - \mathbf{z}'_i \|_2^2,
\end{equation}
where $f'(X'_{i}; \theta')$ is the model's predicted coordinate vector for sample $X'_{i}$. This approach allows the model to leverage its learned spatial-temporal features to perform precise localization.}

\begin{table}[t]
\centering

\renewcommand{\arraystretch}{1.4} 
\begin{tabular}{|p{1.1cm}|p{1.4cm}|p{1cm}|p{2cm}|p{1.1cm}|}
\hline
\textbf{Dataset \& Data Type} & \textbf{Technologies} & \textbf{Sampling Rate} & \textbf{Center Frequencies} & \textbf{Capture Device}  \\ \hline \hline
\makecell{\cite{s17092081} \\ short-range} & LTE, Wi-Fi, DVB-T & 10 MHz & 806 MHz (LTE), 2412/5540 MHz (Wi-Fi), 482 MHz (DVB-T) & Anritsu MS 2690A\\ \hline
\makecell{\cite{9348566} \\ long-range} & Sigfox, LoRa, IEEE 802.15.4g, IEEE 802.11ah & 1 MHz & 863.0 - 868.4 MHz (Sub-GHz) & RTL-SDR dongles\\ \hline
\makecell{\cite{b82jsy5723} \\ short-range} & Wi-Fi (IEEE 802.11ax), LTE, and 5G & 20 MHz & 5.825 GHz (Wi-Fi), 1.8425 GHz (LTE), 628 MHz (5G-NR)  & USRP B210 SDR \\ \hline
\makecell{\Rev{\cite{10195942}} \\ \Rev{UWB CIR}} & \Rev{UWB (LOS/NLOS), Ranging error correction} & \Rev{64 MHz PRF ($\sim$1 ns resolution)} & \Rev{Ch. 1-3 (3.5-4.5 GHz), Ch. 5 (6.5 GHz)} & \Rev{Qorvo DW1000 (Wi-PoS \&  DWM1001 -DEV)} \\ \hline
\end{tabular}
\caption{Overview of the datasets, showcasing the diversity of technologies, sampling rate, and capturing devices used for pre-training.}
\label{tab:datasets}
\end{table}

\subsection{Dataset Description}
The datasets used in this study encompass a much broader range of wireless conditions, including more technologies and capturing conditions than prior studies. We consider two types of technologies: long-range and short-range, each with varying wireless technologies, sampling rates, center frequencies, and different capturing devices.  

\Rev{For the short-range datasets, we consider two publicly accessible datasets.
\cite{s17092081}, which includes LTE, Wi-Fi, and DVB-T signals captured at a 10 MHz sampling rate using an Anritsu MS 2690A device in different locations in Ghent, Belgium. To ensure a diversity of capturing conditions, the measurements were performed at three distinct locations (north, east, and west sides) inside a large 12×80 m office building. The dataset was compiled from measurements of three different wireless technologies under distinct environmental and traffic conditions. The Wi-Fi signal, operating at 5540 MHz, was captured inside an active office environment with two access points. Measurement points were intentionally varied in distance from the transmitters to create a diverse range of signal strengths. In contrast, the LTE (806 MHz) and DVB-T (482 MHz) signals were recorded in an outdoor-to-indoor scenario, originating from a nearby commercial cell tower and a local TV broadcast tower, respectively.
The captured signals also feature authentic background traffic. The Wi-Fi and LTE measurements include live, uncontrolled data from active users, while the DVB-T signal consists of a continuous, stable broadcast stream without the bursty traffic typical of data networks.}

\Rev{Another short-range dataset from \cite{b82jsy5723} provides Wi-Fi (IEEE 802.11ax), LTE, and 5G-NR signals. These were sampled at 20 MHz using a USRP B200mini-i SDR to capture high-frequency technologies at center frequencies of 5.825 GHz (Wi-Fi), 1.8425 GHz (LTE), and 628 MHz (5G-NR), respectively. All captures were performed over the air with active data traffic to better represent real-world signal characteristics. The Wi-Fi and LTE signals were generated within controlled test-beds. These environments used dedicated access points and client devices to create active, known data transmissions. Specific antennas were employed for these captures: a VERT2450 omni-directional antenna (3 dBi gain) for Wi-Fi and an ANT-LTE-WS-SMA dipole antenna (5.9 dBi gain) for LTE.
In contrast, the 5G-NR signal was captured from a live, public commercial mobile network in an uncontrolled outdoor environment.}

\Rev{For the long-range datasets, we consider the dataset from \cite{9348566}, which focuses on sub-GHz technologies, including Sigfox, LoRa, IEEE 802.15.4g, and IEEE 802.11ah, with a sampling rate of 1 MHz and RTL-SDR dongles as capture devices operating in parallel to cover the 868 MHz band. The data was generated in a lab-style test-bed environment using various development boards as transmitters. }

\begin{figure}[h]
\centering
\includegraphics[width=0.53\textwidth]{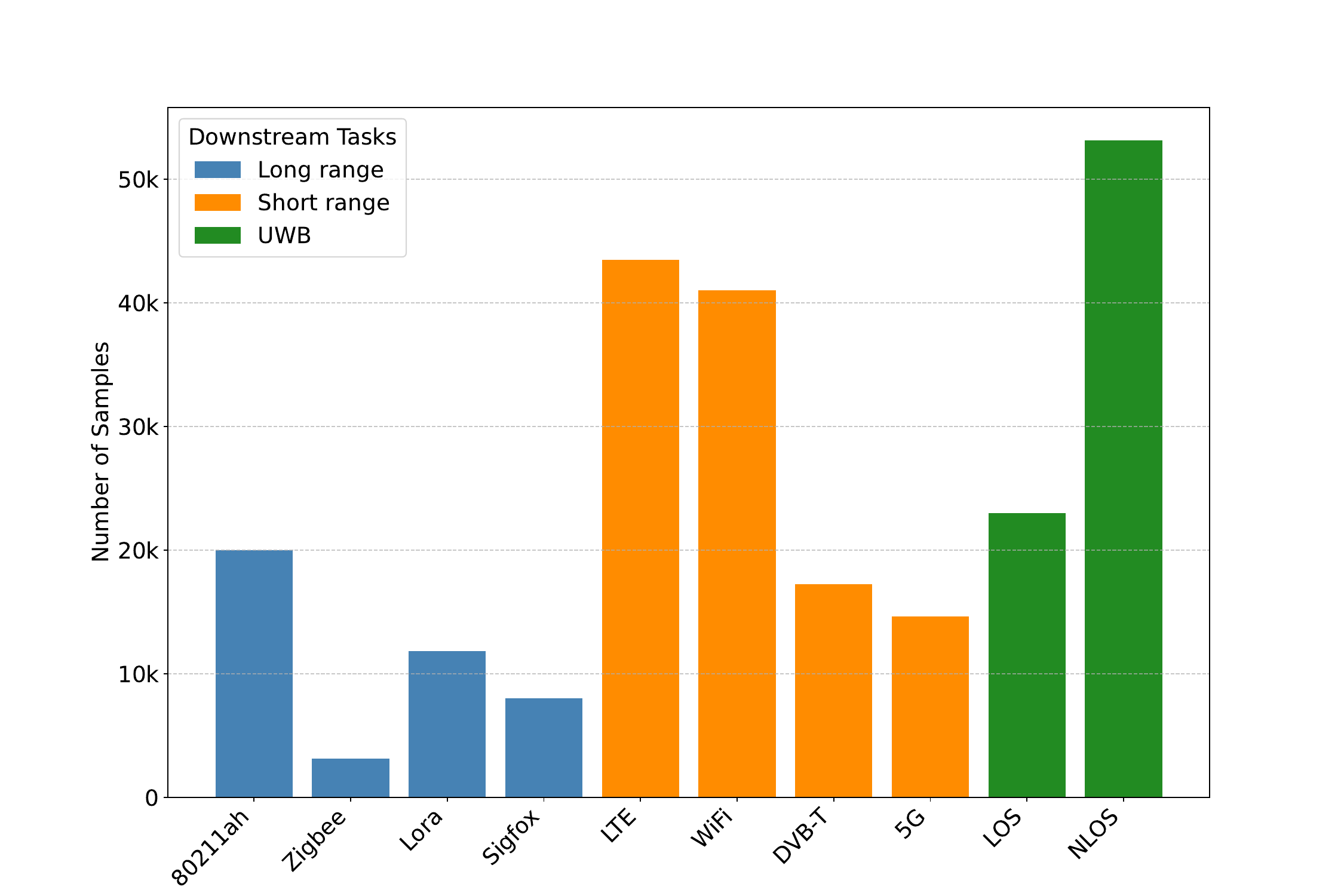}
\caption{\Rev{Distribution of the dataset used for the proposed foundation model.}}
\label{fig:sample_stats}
\end{figure}

\Rev{To evaluate \ac{los} and \ac{nlos} detection, we utilize the comprehensive UWB dataset from \cite{10195942}. This dataset comprises over 80,000 \ac{cir} measurements gathered from three distinct environments: a large industrial warehouse (21,085 samples), a multi-room office (44,894 samples), and a university building (15,208 samples). The data was collected using Qorvo DW1000-based transceivers, specifically the Wi-PoS and DWM1001-DEV platforms, ensuring a variety of hardware configurations. Each measurement is labeled as LOS or NLOS, providing a robust basis for classification tasks. In addition to classification labels, the dataset also includes the ground truth and measured ranges for each sample, enabling the development of models for ranging error correction.

A summary of the datasets is provided in Table~\ref{tab:datasets}. Furthermore, to provide a more detailed statistical overview, Figure~\ref{fig:sample_stats} illustrates the total number of samples available for each distinct technology across all combined datasets. In summary, our dataset contains ten distinct classes, captured with varying sampling rates, at diverse locations, and using different devices, resulting in a broad and more realistic representation of wireless environments compared to existing datasets.}
\section{Methodology} \label{sec:method}
This section introduces our proposed method, providing an overview of its structure and functionality. In addition, we briefly describe the baseline methods used for comparison. Finally, we present the data preparation and implementation details, highlighting the main steps and technical considerations involved in the experimental setup.

\subsection{Proposed Model for Wireless Technology Recognition}

We present an optimized model inspired by the Vanilla Transformer \cite{NIPS2017_3f5ee243} and PatchTST \cite{nie2023timeseriesworth64} architecture for efficient long-term time series analysis in the context of \ac{wtr} \Rev{and localization}. This model employs a combination of patch-based segmentation, channel independence, and attention mechanisms, effectively capturing local and global dependencies while minimizing computational costs.

\subsubsection{Model Architecture Overview} \label{sec:model_arch_overview}
\begin{figure}[t]
\centering
\includegraphics[width=0.49\textwidth]{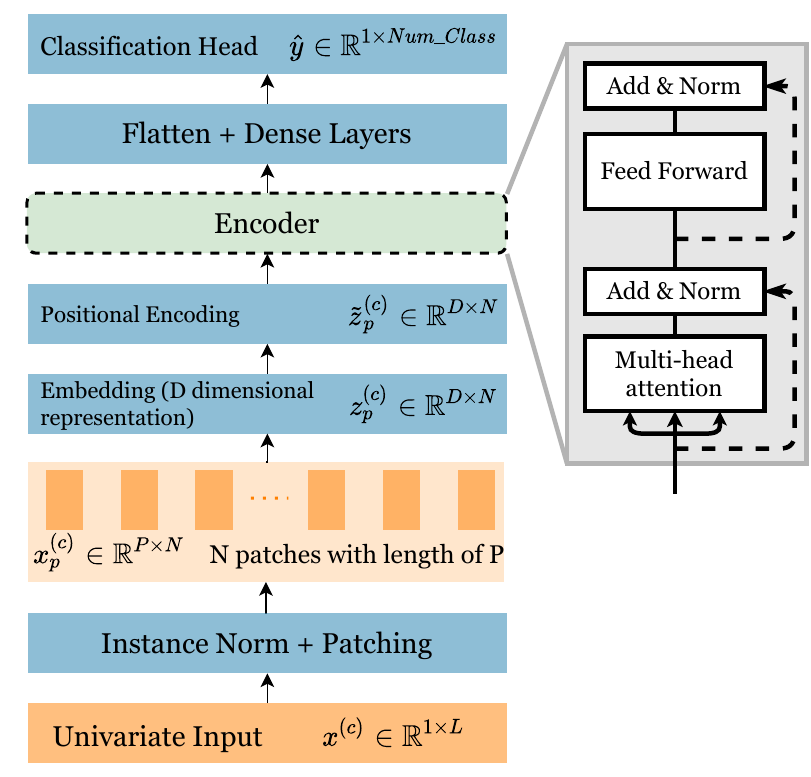}
\caption{\Rev{Proposed Transformer model with patching for wireless technology classification.}}
\label{PatchTST:model}
\end{figure}

The proposed model processes each time series independently, applying a patch-based segmentation approach followed by a Transformer encoder to capture long-term dependencies as depicted in Fig.~\ref{PatchTST:model}. \gls{iq} \Rev{and \ac{cir} samples have real and imaginary parts that} are known as two channels of data \( x = (x^{(r)}, x^{(i)}) \) and are treated independently, without cross-channel interaction, preserving the temporal structure within each channel. This channel-wise independence allows the model to handle large multivariate datasets more efficiently by reducing the complexity involved in attention calculations.

\begin{figure}[t]
\centering
\includegraphics[width=0.45\textwidth]{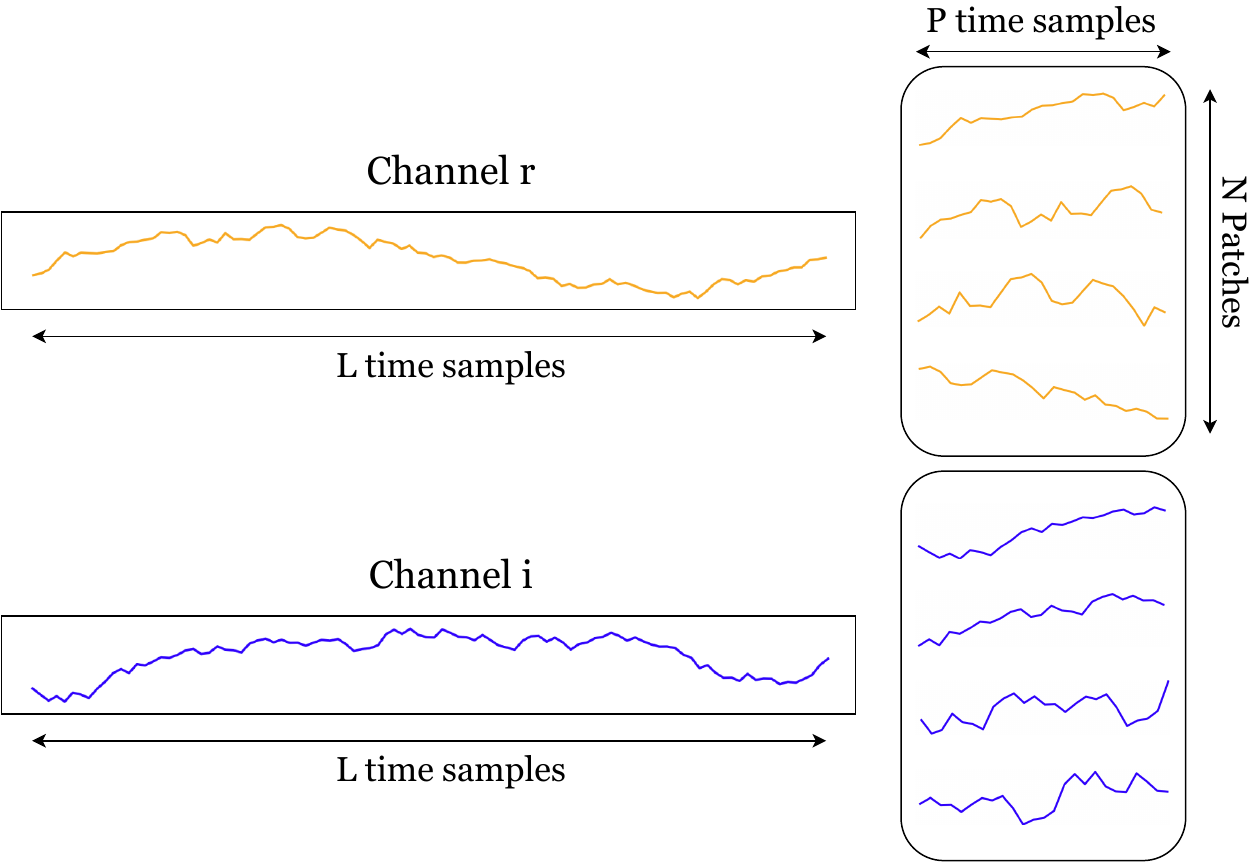}
\caption{Illustration of the patching process applied to \ac{iq} timeseries data. Each channel (r and i) consists of $L$ time samples and is independently divided into $N$ overlapping patches, each of length $P$. These patches are then used as input tokens for the transformer-based model.}
\label{Patching}
\end{figure}

\subsubsection{Patching Strategy}

To address the issue of high computational cost for long time sequences, we segment each time series channel into overlapping patches as illustrated in Fig.~\ref{Patching}. Given a time series \( x^{(r)} \) or \( x^{(i)} \) of length \( L \), each channel is divided into patches of length \( P \) with a stride \( S \). This segmentation reduces the input sequence length from \( L \) to approximately \( \frac{L}{S} \), significantly lowering the computational complexity of the model. A smaller stride increases patch overlap, allowing smoother transitions between patches but also increasing computational load.

For a single channel (either \( r \) or \( i \)), let the segmented patches be represented by:

\begin{multline} \label{eq:single_ch_patch}
x^{(c)}_p = \left\{ x^{(c)}_{t}, x^{(c)}_{t+1}, \dots, x^{(c)}_{t+P-1} \right\} \quad \\
\forall t = \{1, S, 2S, \dots, L-P+1\},
\end{multline}

\noindent where \( c \in \{r, i\} \). These patches \( x^{(c)}_p \in \mathbb{R}^{P \times N} \), where \( N = \frac{L}{S} \), serve as the primary input tokens to the Transformer.  The patching reduces the effective sequence length from \( L \) to \( N \), which lowers the complexity of the attention mechanism from \( \mathcal{O}(L^2) \) to \( \mathcal{O}(N^2) \).

Each patch \( x^{(c)}_p \) is then projected into a latent space using a linear projection matrix \( W_p \in \mathbb{R}^{D \times P} \), mapping each patch to a \( D \)-dimensional representation \cite{NIPS2017_3f5ee243}:
\begin{equation} \label{eq:embed}
z^{(c)}_p = W_p x^{(c)}_p + b_p,
\end{equation} 

\noindent where \( b_p \in \mathbb{R}^D \) is the bias term. This transformation effectively captures the local temporal patterns within each patch, and a positional encoding \( E_{pos} \in \mathbb{R}^{D \times N} \) is an addition to the projected patches to preserve the temporal sequence order since the Transformer architecture lacks inherent order awareness:
\begin{equation} \label{eq:positoin}
\tilde{z}^{(c)}_p = \text{Dropout}(z^{(c)}_p + E_{pos}),
\end{equation} 

\noindent where dropout is applied to the summed embedding to prevent overfitting and improve generalization.

\subsubsection{Multi-Head Self-Attention Mechanism}

The positionally encoded patch representations \( \tilde{z}^{(c)}_p \), derived from the input patches of channel \( c \), are passed to the Transformer encoder’s multi-head self-attention module \cite{NIPS2017_3f5ee243}. This mechanism is designed to capture both local and global dependencies across the sequence.

For each attention head \( h = \{1, \dots, H \} \), where \(H\) is the number of attention heads, the attention mechanism computes query, key, and value matrices \( Q_h \), \( K_h \), and \( V_h \) as follows:

\begin{align} \label{eq:attention_mech}
Q_h = W_Q^{(h)} \tilde{z}^{(c)}_p, \\ \quad K_h = W_K^{(h)} \tilde{z}^{(c)}_p, \\ \quad V_h = W_V^{(h)} \tilde{z}^{(c)}_p,
\end{align}

\noindent here $\tilde{z}^{(c)}_p$ represents the input sequence of positionally encoded patch embeddings from channel $c$, and \( W_Q^{(h)}, W_K^{(h)}, W_V^{(h)} \in \mathbb{R}^{D \times d_k} \) are learnable weight matrices, and \( d_k = D / H \) is the dimension of each attention head.

The attention scores are computed as:

\begin{equation} \label{eq:attention_score}
\text{Attention}(Q_h, K_h, V_h) = \text{Dropout}(\text{Softmax}\left(\frac{Q_h K_h^\top}{\sqrt{d_k}}\right) V_h).
\end{equation}

\noindent The scaling factor \( \sqrt{d_k} \) is used to avoid large dot products, which can lead to vanishing gradients in the softmax function.

The outputs from all attention heads are concatenated along the feature dimension and linearly transformed using a shared output projection matrix \( W_O \in \mathbb{R}^{D \times D} \):

\begin{equation}
\hat{z}^{(c)}_p = W_O \left[ \text{Attention}_1, \dots, \text{Attention}_H \right].
\end{equation}

\noindent where \( \hat{z}^{(c)}_p \in \mathbb{R}^{D \times N} \) is the final attended patch representation. This multi-head setup allows the model to attend to information from different representation subspaces simultaneously, improving its capacity to model complex temporal dependencies.

This mechanism enables the model to assign greater importance to the most relevant patches, enhancing its ability to generalize across diverse wireless signals while maintaining interpretability through the learned attention patterns.

\subsubsection{Classification Head}
The encoded features from both channels are concatenated and flattened:

\begin{equation}
f = \text{Flatten} \left( \left[ \hat{z}^{(r)}_p \| \hat{z}^{(i)}_p \right] \right),
\label{eq:flatten}
\end{equation}

\noindent which is then passed through a two-layer fully connected network:

\begin{equation} \label{eq:classification}
\hat{y} = W_{c1} \, \sigma(W_{c2} f + b_{c2}) + b_{c1},
\end{equation}

\noindent where \( \sigma \) is the ReLU activation, and \( W_{c1}, W_{c2}, b_{c1}, b_{c2} \) are learnable parameters. The output \( \hat{y} \in \mathbb{R}^{C} \) represents class logits, which are passed through a softmax during inference to generate the predicted class probabilities.

\Rev{
\subsubsection{Regression Head}
Similar to the classification head, the encoded features from both channels are first concatenated and flattened as in eq~\ref{eq:flatten}. The combined feature vector is then processed by a two-layer fully connected network:

\begin{equation} \label{eq:regression} \hat{\mathbf{z}} = W_{r1} , \sigma(W_{r2} f + b_{r2}) + b_{r1}, 
\end{equation}

\noindent where \( \sigma \) is the ReLU activation, and \( W_{r1}, W_{r2}, b_{r1}, b_{r2} \) are the learnable parameters of the regression head. The output \( \hat{\mathbf{z}} \in \mathbb{R}^{d} \) is the predicted continuous vector, where \( d \) is the dimension of the output.}

\subsection{Baseline Models (WTR)}
We consider four baseline models to evaluate the performance of our proposed method in classifying wireless communication technologies across heterogeneous classes: 802.11ah, Zigbee, LoRa, Sigfox, LTE, WiFi, DVB-T, and 5G.

\subsubsection{CNN} \label{cnn:baseline}
We implement \ac{cnn}  model as a baseline based on a research paper in \ac{wtr} \cite{8824856}. The architecture is composed of three convolutional layers, followed by two fully connected layers. Each convolutional layer uses ReLU activation, followed by batch normalization, max-pooling, and dropout to reduce overfitting and enhance feature extraction. The fully connected layers further process the flattened features, and the final layer uses a softmax activation for multi-class classification.

This model was originally designed and optimized for a small number of wireless technology classes. In our study, we include this model as a baseline to investigate how such specifically tailored models perform when scaled to more complex classification tasks involving a larger number of heterogeneous technologies.

\subsubsection{Autoencoder + CNN}  
We employ a convolutional autoencoder to learn a compact latent representation of wireless signal data in an self-supervised manner \cite{8935690}. The architecture consists of an encoder with two convolutional layers, followed by a fully connected layer. The decoder mirrors this structure using transposed convolutional layers to reconstruct the input from the latent space. Batch normalization and ReLU activations are applied throughout the network to stabilize training and improve learning dynamics. Given that the input consists of 4096 time samples, the dimensionality of the latent space in the autoencoder is set to 1024.

After training the autoencoder using only unlabeled examples, the encoder is frozen and reused as a generic feature extractor. The latent representation serves as input to a CNN model. We used the model explained in \ref{cnn:baseline}.

\subsubsection{LSTM}  
\ac{rnn}, particularly \ac{lstm} networks, are widely used for modeling sequential data due to their ability to capture dependencies in time series. Inspired by the architecture presented in \cite{9171706}, we process the input \ac{iq} samples using three stacked LSTM layers, followed by fully connected layers and a softmax output for classification. Dropout regularization is applied between layers to prevent overfitting, and ReLU activations are used in the dense layers to ensure non-linearity.

While LSTMs are theoretically well-suited for modeling signal dynamics, our experiments show that this architecture is significantly slower than other models in both pre-training and adaptation. The sequential nature of LSTM computations prevents efficient parallelization, which becomes a bottleneck when dealing with large datasets or high-throughput applications.

\begin{figure}[t]
\centering
\includegraphics[width=0.7\columnwidth]{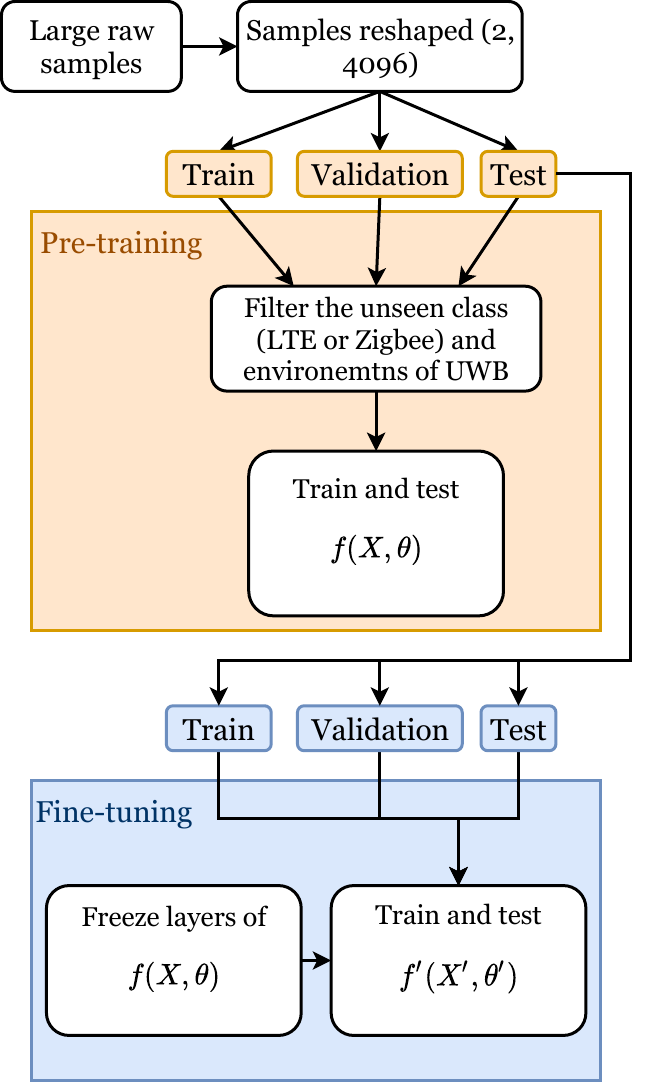}
\caption{\Rev{Flow-chart depicting the pre-training and fine-tuning strategy}}
\label{fig:flowchart}
\end{figure}

\subsubsection{Transformer-based Models}  
We also include two Transformer-based models in our comparison: the vanilla Transformer and the iTransformer. We do not propose these models, but they serve as state-of-the-art baselines for comparison with our method. These models represent different tokenization strategies for timeseries data: (1) individual time samples as tokens (vanilla Transformer), (2) patches of time samples (our method), and (3) entire channels as tokens (iTransformer). By comparing across these architectures, we aim to highlight how tokenization granularity and model complexity influence performance.

\textbf{Vanilla Transformer:} The vanilla Transformer \cite{NIPS2017_3f5ee243}, originally developed for language modeling, has been adapted for timeseries classification by treating each time step as a separate token. This results in long input sequences and leads to quadratic complexity \( \mathcal{O}(N^2) \) in the attention mechanism, significantly increasing computational cost for high-resolution signals. The model applies multi-head self-attention followed by feed-forward layers, layer normalization, and residual connections. While effective in learning sequence dependencies, its deep architecture and token-level granularity make it computationally heavy.

\textbf{iTransformer:} The iTransformer \cite{liu2024itransformer} is a more efficient alternative that keeps the core Transformer components unchanged but inverts the input dimensions. Instead of treating time steps as tokens, it treats the channels (or variates) of the time series as tokens by embedding all time points of each channel. The attention mechanism is then applied across these channel tokens to capture multivariate correlations. The feed-forward layers are applied independently to each token to learn non-linear features. This simple yet effective design makes the model less complex while also improving generalization and allowing flexible use of different lookback windows.

\Rev{
\subsection{Baseline Model (UWB)}
To evaluate the performance of our proposed method in \ac{los} detection and localization error correction, we use a baseline inspired by the work of \cite{10195942}. This study, focused on UWB ranging error correction and \ac{los} signal classification, proposes two parallel architectures: a feature-based \ac{dnn} that operates on 12 features extracted from the UWB transceiver, and a \ac{cnn} that learns directly from raw \ac{cir} data. The core of their work is an automated Transfer Learning (TL) framework that uses Bayesian optimization to adapt a pre-trained model to a new, unseen environment with smaller samples. We include this baseline to evaluate performance on a practical task where models must overcome data scarcity and generalize across different physical environments and hardware configurations, a common challenge in wireless sensing applications.}

\subsection{Data Preparation for pre-training and fine-tuning}

To evaluate the generalization capabilities of our models, we designed structured data handling to simulate unseen class scenarios, as depicted in Fig.~\ref{fig:flowchart}. First, a large dataset containing \ac{iq} \Rev{and \ac{cir}} samples was prepared. Each sample was reshaped into a uniform shape of (2, 4096), where '2' represents the real and imaginary channels, as introduced in Section IV.\ref{sec:model_arch_overview}. This reshaping ensures that all data instances maintain a consistent input format, simplifying the pre-training and adaptation process.

To evaluate the model's generalization capability in the presence of an unseen class \Rev{or environment} during pre-training, we excluded specific classes \Rev{for WTR and environments for UWB} from the dataset. In particular, either Zigbee (the class with the fewest samples) or LTE (the class with the most samples) was designated as the unseen class. Zigbee is also a class with a single sampling rate and collected from a single source, while LTE includes two sampling rates and data gathered from multiple sources. Following this exclusion, the dataset was divided into three subsets: training, validation, and test. The model was trained on the training set, tuned using the validation set, and finally evaluated on the test set. Importantly, the test set was kept completely separate from the training data to ensure an unbiased assessment of the model's performance.

The unseen class was included in the test data in the fine-tuning phase. Now containing the previously unseen class, the extended test dataset was again split into new train, validation, and test sets. During fine-tuning, layers of the models were frozen, with only the dense output layer being trained. In our simulations, our proposed model uses the parameters mentioned in Table.~\ref{tab:parameters}.

\begin{table}[t]
\centering

\renewcommand{\arraystretch}{1.1} 
\begin{tabular}{|p{1.2cm}|p{3.5cm}|p{1.2cm}|}
\hline
\textbf{Parameter} & \textbf{Description}  & \textbf{value}  \\ \hline \hline

$L$ & Input sequence length & 4096\\ \hline

$S$ &  Patch stride & 128 \\ \hline 

$P$ &  Patch size & 128 \\ \hline 

$c$ & Number of input channels  & 2 \\ \hline 

$D$ &Dimensionality of the encoder layers  & 128 \\ \hline 

 -   & Number of encoder layers & 4 \\ \hline

\end{tabular}
\caption{Summary of key parameters and values}
\label{tab:parameters}
\end{table}

\subsection{Implementation} \label{sec:implementation}
We implemented the models using PyTorch and TensorFlow, leveraging a single Tesla V100-SXM3-32GB GPU for both training and fine-tuning processes in the IDLab GPULab \cite{gpulab_idlab}. For the baseline models and our framework, we constructed custom HDF5 datasets to handle the large-scale, multivariate wireless communication timeseries data efficiently. The HDF5 structure allows for optimized storage, retrieval, and scalability, ensuring efficient data loading during training and evaluation. We configured data loaders with a batch size of 64. For pre-training, we used the AdamW optimizer with a learning rate of $1\times 10^{-4}$ and a weight decay of $1\times 10^{-4}$, applied to all model parameters. For fine-tuning, we used AdamW with a lower learning rate of $1\times 10^{-5}$ and a weight decay of $1\times 10^{-5}$. In both phases, we employed a ReduceLROnPlateau scheduler that monitors the validation loss and reduces the learning rate by a factor of 0.7 if no improvement is observed for 3 consecutive epochs.

\section{Results and Discussion} \label{sec:results}

This section presents an evaluation of our proposed method and the baseline models under various scenarios. We begin by analyzing the performance of models during the supervised training stage. Next, we discuss the \ac{wtr} fine-tuning results, where the models are adapted to recognize all classes, even when some were not present during pre-training \Rev{and UWB fine-tuning results, where our model is adapted for regression and classification tasks considering unseen environment scenarios and compared with the baseline model}. Finally, we provide a complexity analysis to assess the computational efficiency of our method in comparison to existing approaches.
\Rev{
\subsection{WTR Supervised Results}}
In this section, we present the results of supervised pre-training under two main scenarios:
\begin{enumerate}
    \item \textbf{8-class scenario:} assumes that all eight wireless technologies are available during training, which works as the baseline to compare with fine-tuning results.
    \item \textbf{7-class scenario:} introduces a constraint where one technology is excluded during training to simulate an unseen class. This may be due to the unavailability of class-specific samples or an insufficient number of pre-training samples.
\end{enumerate}

\begin{table}[t]
    \centering
    \caption{Supervised Pre-training Results: \Rev{These supervised results are presented to establish a performance baseline for evaluating the fine-tuning results of our foundation model.}}
    \begin{subtable}[t]{0.45\textwidth}
        \centering
        \caption{8-class scenario (all classes seen during training)}
        \label{8class:pretrain}
        \begin{tabular}{lc}
            \toprule
            \textbf{Model} & \textbf{Accuracy} \\
            \midrule
            CNN                    & 54.89\% \\
            LSTM                   & \Rev{75.62}\% \\
            Autoencoder + CNN      & 78.01\% \\
            iTransformer           & 74.19\% \\
            Vanilla Transformer    & 65.32\%  \\
            \midrule
            Our method               & \textbf{99.47\%} \\
            \bottomrule
            \hspace{0.03\textwidth}
        \end{tabular}
    \end{subtable}
    \hfill
    \begin{subtable}[t]{0.45\textwidth}
        \centering
        \caption{7-class scenario with excluded technologies}
        \label{7class:pretrain}
        \begin{tabular}{lcc}
            \toprule
            \textbf{Model} & \textbf{LTE Excluded } & \textbf{Zigbee Excluded} \\
            \midrule
            CNN                 & 63.97\% & 57.77\% \\
            LSTM                & \Rev{72.15}\% & \Rev{73.82}\% \\
            Autoencoder + CNN   & 83.53\% & 79.54\% \\
            iTransformer   & 79.84\% & 74.43\% \\
            Vanilla Transformer  & 74.76\% & 66.42\% \\
            \midrule
            Our method               & \textbf{99.69\%} & \textbf{99.66\%} \\
            \bottomrule
        \end{tabular}
    \end{subtable}
    \label{tab:supervised_results}
\end{table}

\begin{table*}[t]
    \caption{Fine-tuning Results on Different Downstream Tasks}
    \label{tab:finetuning_results_aligned}
    \centering
    \begin{tabular}{lcccc}
        \toprule
        \textbf{Model} & \multicolumn{2}{c}{\textbf{LTE Excluded}} & \multicolumn{2}{c}{\textbf{Zigbee Excluded}} \\
        \cmidrule(lr){2-3} \cmidrule(lr){4-5}
        & 8-class & short-range & 8-class & long-range \\
        \midrule
        CNN                      & 39.50\% & 62.38\% & 43.45\% & 46.08\% \\
        LSTM                     & \Rev{57.41}\% & \Rev{52.43}\% & \Rev{69.50}\% & \Rev{81.36}\% \\
        Autoencoder + CNN        & 71.60\% & 61.87\% & 79.36\% & 97.45\% \\
        iTransformer & 59.10\% & 52.79\% & 74.43\% & 91.10\% \\
        \midrule
        Our method (Supervised)              & 62.79\% & 86.4\% & \textbf{98.3\%} & 99.52\% \\
        Our method (Self-supervised) & \textbf{\Rev{93.42}\%} & \textbf{\Rev{91.14}\%} & \Rev{96.84}\% & \textbf{\Rev{99.99}\%} \\
        \bottomrule
    \end{tabular}
\end{table*}

Table~\ref{8class:pretrain} presents the classification results for the 8-class scenario, where all wireless technologies are included during training. Traditional models such as CNN and LSTM achieve accuracies of 54.89\% and \Rev{75.62}\%, respectively. Their lower performance is due to the increased number of classes and the presence of heterogeneous sampling rates, which challenge their ability to generalize across diverse IQ data. Incorporating an autoencoder with CNN improves the performance to 78.01\%. Among Transformer-based models, the vanilla Transformer and iTransformer obtain 71.03\% and 74.19\% accuracies, respectively. In contrast, our proposed method achieves a significantly higher accuracy of 99.47\%, demonstrating its strong learning capability in a supervised setting with heterogeneous input data. These results highlight that classical models and even standard Transformer variants struggle when exposed to more diverse classes and varying sampling rates. While the vanilla Transformer treats each time step as a token, and the iTransformer treats each I/Q channel as a token, they fail to capture the underlying structure as effectively as our approach.

Table~\ref{7class:pretrain} presents results for the 7-class scenario, where one class is held out during training. This experiment is designed to evaluate the model’s ability to generalize to unseen technologies in the fine-tuning stage. We investigate two cases: (1) excluding LTE, the most dominant class in terms of sample count and sampling rate diversity, and (2) excluding Zigbee, the least represented class with only one sampling rate.

When either LTE or Zigbee is excluded from the training set, most models exhibit improved performance compared to the full 8-class scenario. For example, CNN achieves accuracies of 63.97\% and 57.77\% when LTE and Zigbee are excluded, respectively. Similarly, LSTM reaches \Rev{72.15}\% accuracy in the LTE-excluded case and \Rev{73.82}\% in the Zigbee-excluded case. Interestingly, LSTM is the only model that shows a performance drop in the \Rev{LTE}-exclusion setup compared to the 8-class case. For all other models, the performance trend aligns with CNN, showing increased accuracy upon exclusion. This improvement is more prominent when excluding LTE, likely because the LTE dataset introduces higher variability due to its different sampling rates, making it harder for traditional models to generalize. \Rev{This observation reinforces that the performance of baseline models is fragile and degrades significantly as task complexity increases, which accounts for the large performance gap noted in the 8-class scenario.} Despite these variations, our proposed method consistently outperforms all baselines, achieving 99.69\% and 99.66\% accuracy in the LTE- and Zigbee-excluded scenarios, respectively. \Rev{These results highlight that our method performs well even when facing eight or seven technologies. The comparison between the two exclusion cases further emphasizes the robustness of our approach, especially in handling variations due to sampling rate.}

These results highlight that our method performs well even when facing eight or seven technologies. The comparison between the two exclusion cases further emphasizes the robustness of our approach, especially in handling variations due to sampling rate.

\subsection{WTR Fine-Tuning Results}

We now assess the model performance during the fine-tuning stage, where a limited number of labeled samples from the previously unseen class are provided. This step reflects a realistic scenario in which a pre-trained model is adapted to recognize new wireless technologies using only a small amount of labeled data. We focus on \textbf{7-class scenario} (two exclusions) earlier: the \textbf{LTE-excluded} and \textbf{Zigbee-excluded} setups.

Table~\ref{tab:finetuning_results_aligned} summarizes the fine-tuning results on downstream tasks using all available data for fine-tuning. \Rev{The performance trends highlight the model's sensitivity to the pre-training data, which results from the differing complexity of the excluded classes. LTE represents a "hard" class due to its diversity and spectral overlap with other signals, while Zigbee is an "easy" class with clean, distinct patterns. This difference explains the results. When the "hard" LTE class is excluded, our self-supervised model shows its strength, achieving 93.42\% on the 8-class task, far surpassing all other models, including our supervised version (62.79\%). This demonstrates that self-supervised learning builds more generalizable representations, which are essential for adapting to a complex, unseen class. Conversely, when the "easy" Zigbee class is excluded, all models are pre-trained on a more complex dataset (containing LTE), providing a robust foundation. Consequently, both our supervised and self-supervised models achieve outstanding accuracy, with the supervised model reaching 98.3\% in the 8-class task and the self-supervised model peaking at 99.99\% in the long-range task. This illustrates that the diversity of the pre-training data is key to achieving high performance.}

These results highlight the strong generalization capability of the self-supervised model, which, despite only receiving labeled data during fine-tuning, performs well across all tasks. Notably, in the LTE-excluded 8-class scenario, the supervised model underperforms (62.79\%) compared to its self-supervised counterpart (93.42\%), indicating the difficulty of adapting to this specific unseen class using limited supervision. This further emphasizes the effectiveness of self-supervised pre-training in enabling the model to better adapt to new, unseen technologies with minimal labeled data. One possible reason for the lower performance in the LTE-excluded setting is that LTE data in the dataset spans multiple sampling rates and shares spectral characteristics with 5G, making it harder for a purely supervised model to learn distinctive features from a few fine-tuning samples.

\begin{figure}[t]
    \centering
    \begin{subfigure}[b]{0.44\textwidth}
        \centering
        \includegraphics[width=\textwidth]{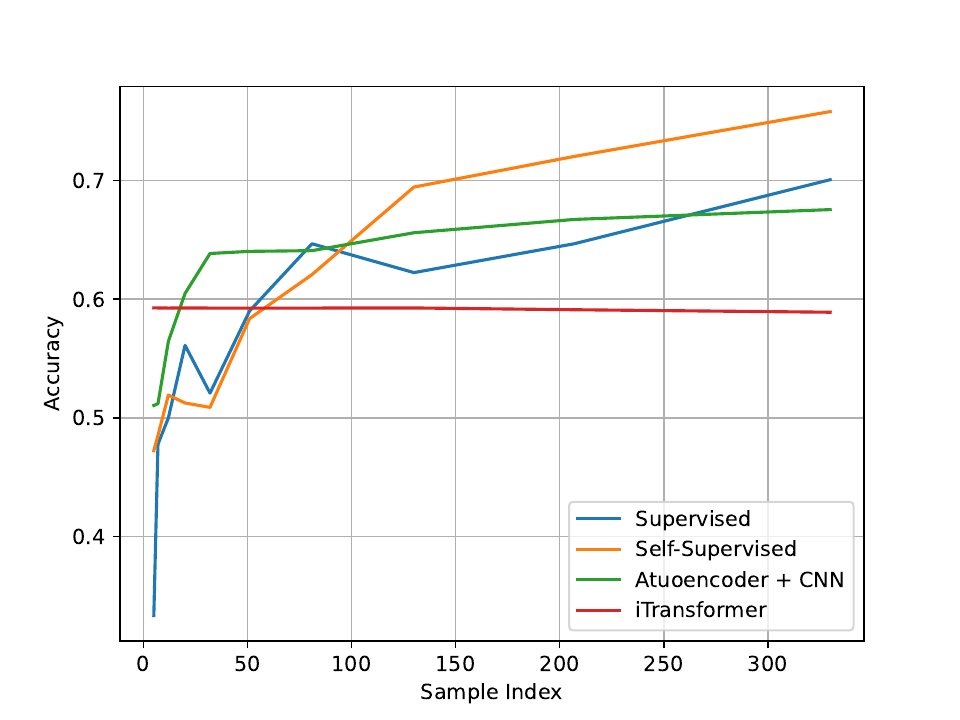}
        \caption{LTE-excluded case.}
        \label{fig:lte_unseen}
    \end{subfigure}
    \begin{subfigure}[b]{0.49\textwidth}
        \centering
        \includegraphics[width=\textwidth]{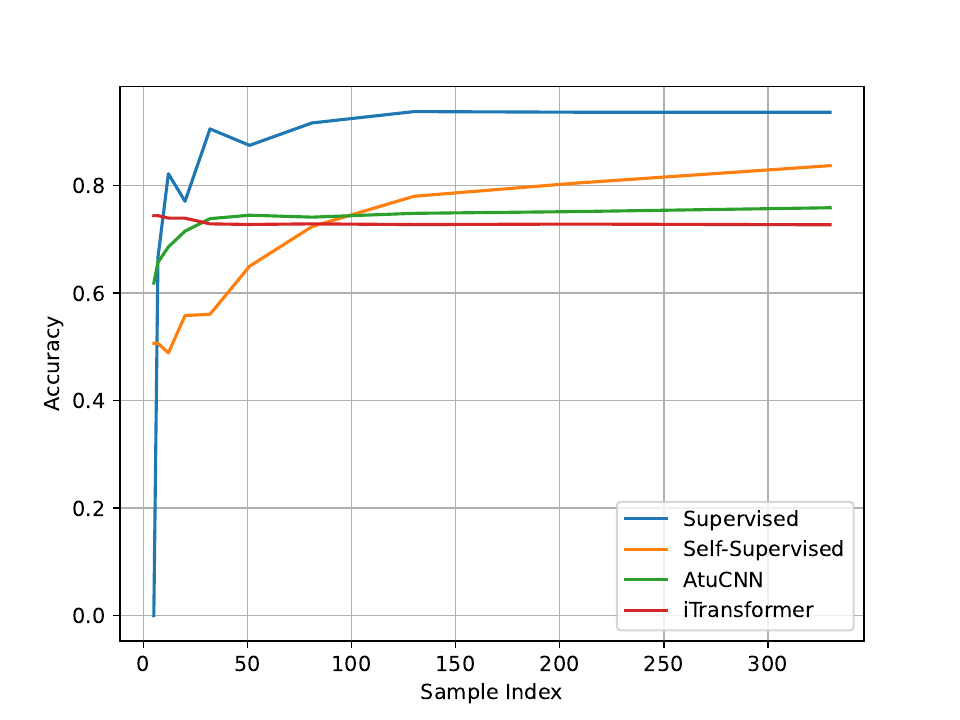}
        \caption{Zigbee-excluded case.}
        \label{fig:zigbee_unseen}
    \end{subfigure}
    \caption{Fine-tuning accuracy when a class was excluded during pre-training. The x-axis shows the number of samples per class used for fine-tuning.}
    \label{fig:unseen_class_tuning}
\end{figure}

Figures~\ref{fig:lte_unseen} and~\ref{fig:zigbee_unseen} illustrate the fine-tuning accuracy of different models on two excluded-class scenarios: LTE and Zigbee. The x-axis indicates the number of samples used for fine-tuning per class, while the y-axis shows the resulting accuracy. \Rev{For these experiments, the model is fine-tuned and evaluated on the complete 8-class scenario.} Among the models compared, we specifically focus on the performance of our supervised and self-supervised solutions in relation to Autoencoder + CNN and iTransformer, which were selected for this analysis due to their superior performance in prior evaluations. In contrast, LSTM and Vanilla Transformer models were excluded from this comparison due to their significantly higher computational demands, which rendered them impractical for large-scale fine-tuning experiments.

\begin{figure}[t]
    \centering
    
    \begin{subfigure}[b]{0.24\textwidth}
        \centering
        \includegraphics[width=\textwidth]{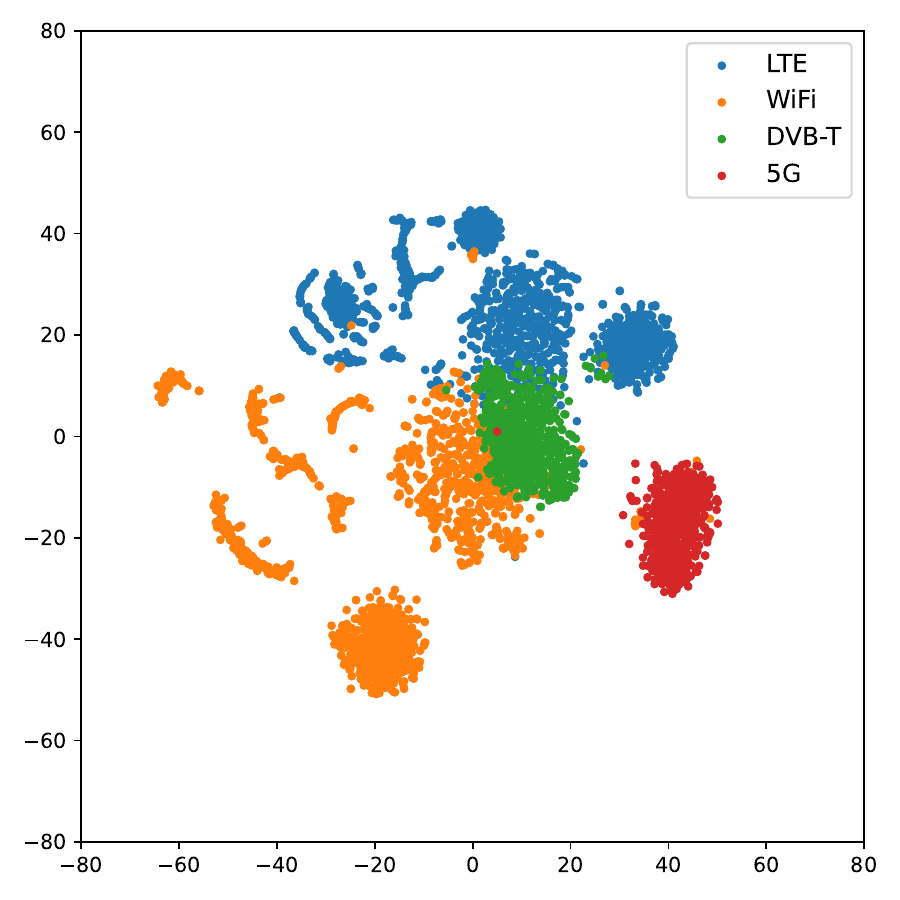}
        \caption{Short-range}
        \label{fig:a}
    \end{subfigure}
    \hfill 
    \begin{subfigure}[b]{0.24\textwidth}
        \centering
        \includegraphics[width=\textwidth]{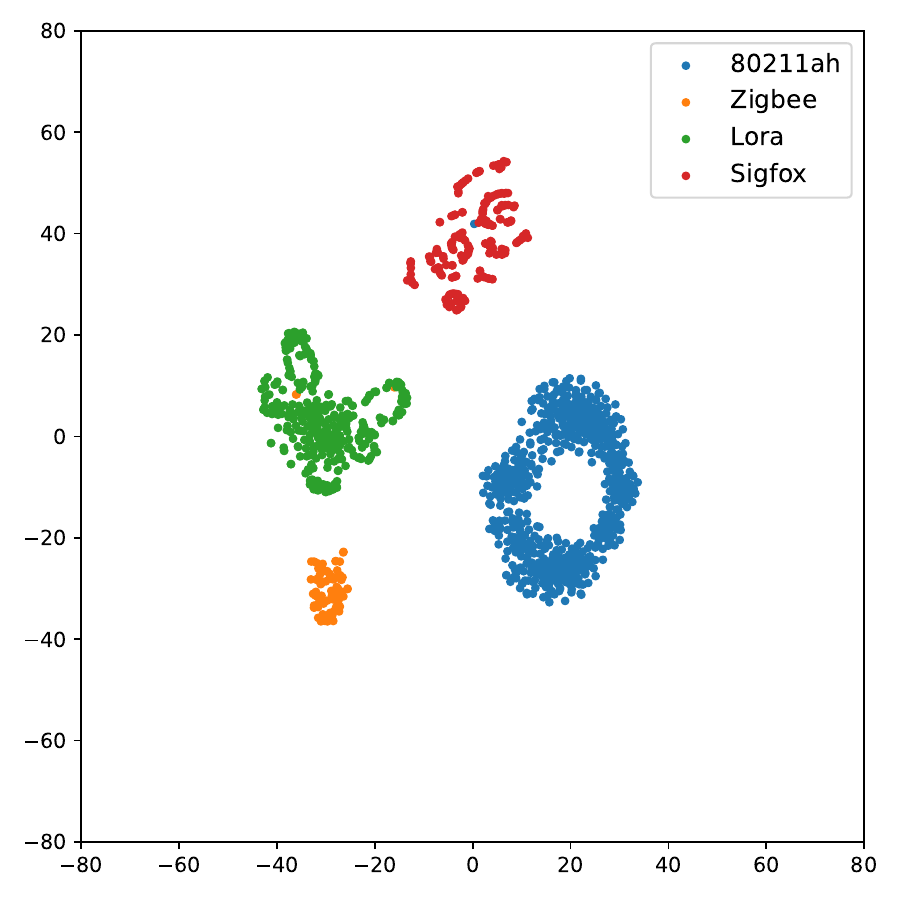}
        \caption{Long-range}
        \label{fig:b}
    \end{subfigure}
    \hfill 
    \begin{subfigure}[b]{0.24\textwidth}
        \centering
        \includegraphics[width=\textwidth]{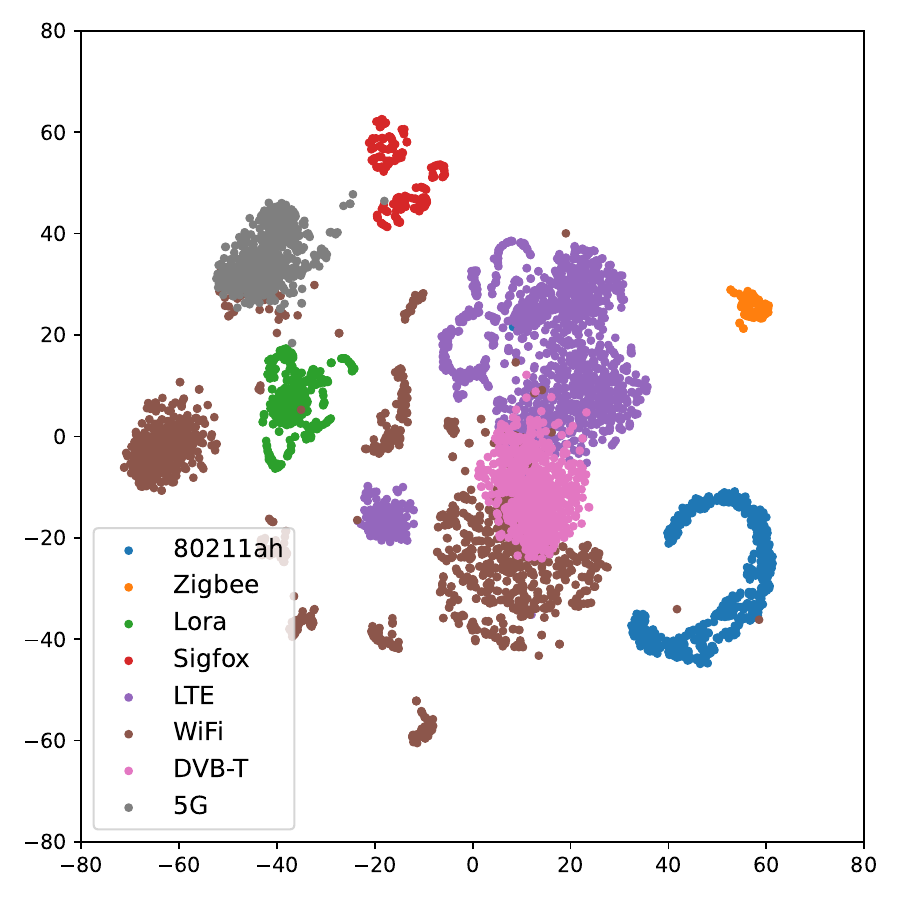}
        \caption{8-class}
        \label{fig:c}
    \end{subfigure}
    \hfill 
    \begin{subfigure}[b]{0.24\textwidth}
        \centering
        \includegraphics[width=\textwidth]{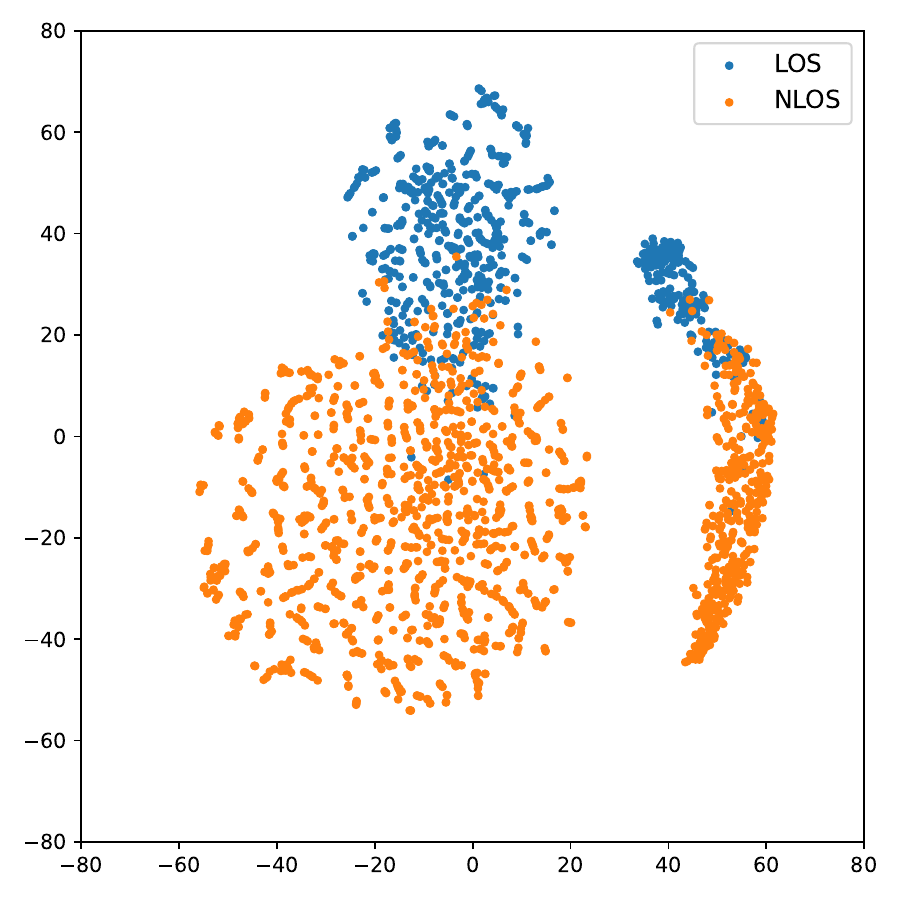} 
        \caption{UWB (N)LOS}
        \label{fig:d}
    \end{subfigure}

    \caption{\Rev{t-SNE visualization of feature embeddings from the self-supervised model (all classes seen) after it was fine-tuned for four distinct downstream tasks. The clear clustering in each plot demonstrates the model's strong ability to learn discriminative features for different classification scenarios.}}
    \label{fig:tsne}
\end{figure}

\begin{figure*}[h]
    \centering

    \begin{subfigure}{\textwidth}
        \includegraphics[width=\textwidth]{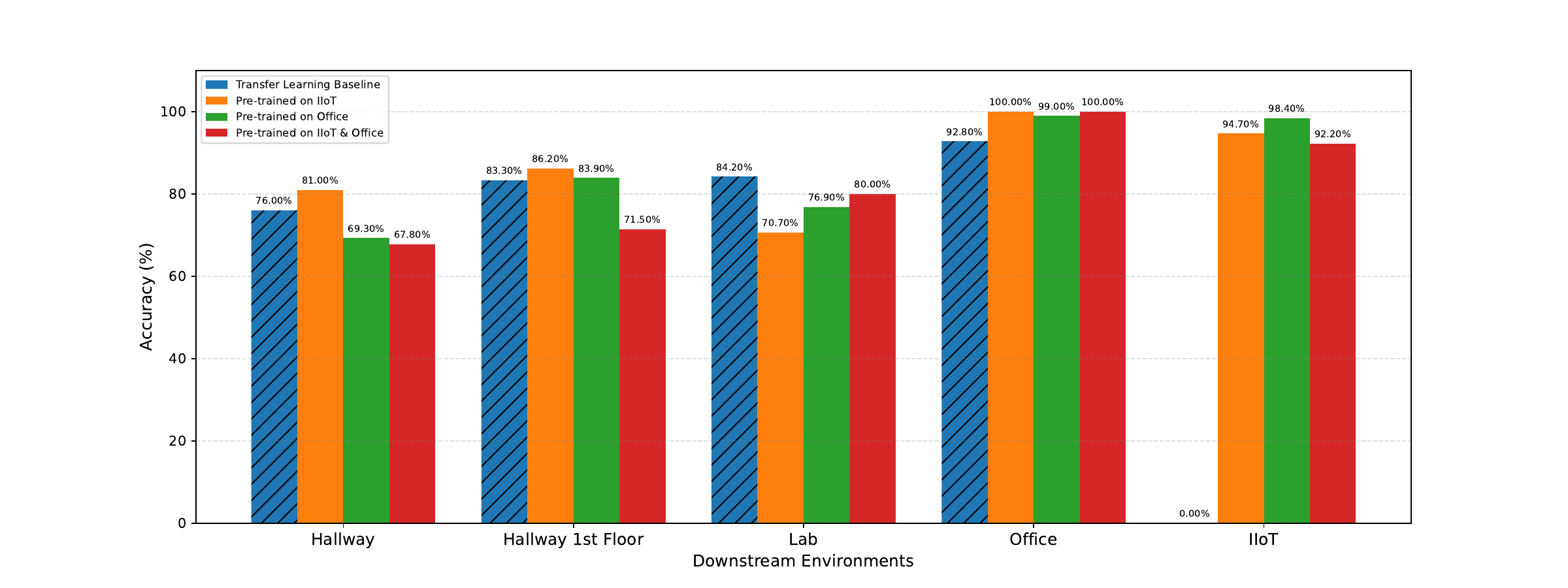}
        \caption{\Rev{(N)LOS classification accuracy.}}
        \label{fig:nlos_env}
    \end{subfigure}

    \begin{subfigure}{\textwidth}
        \includegraphics[width=\textwidth]{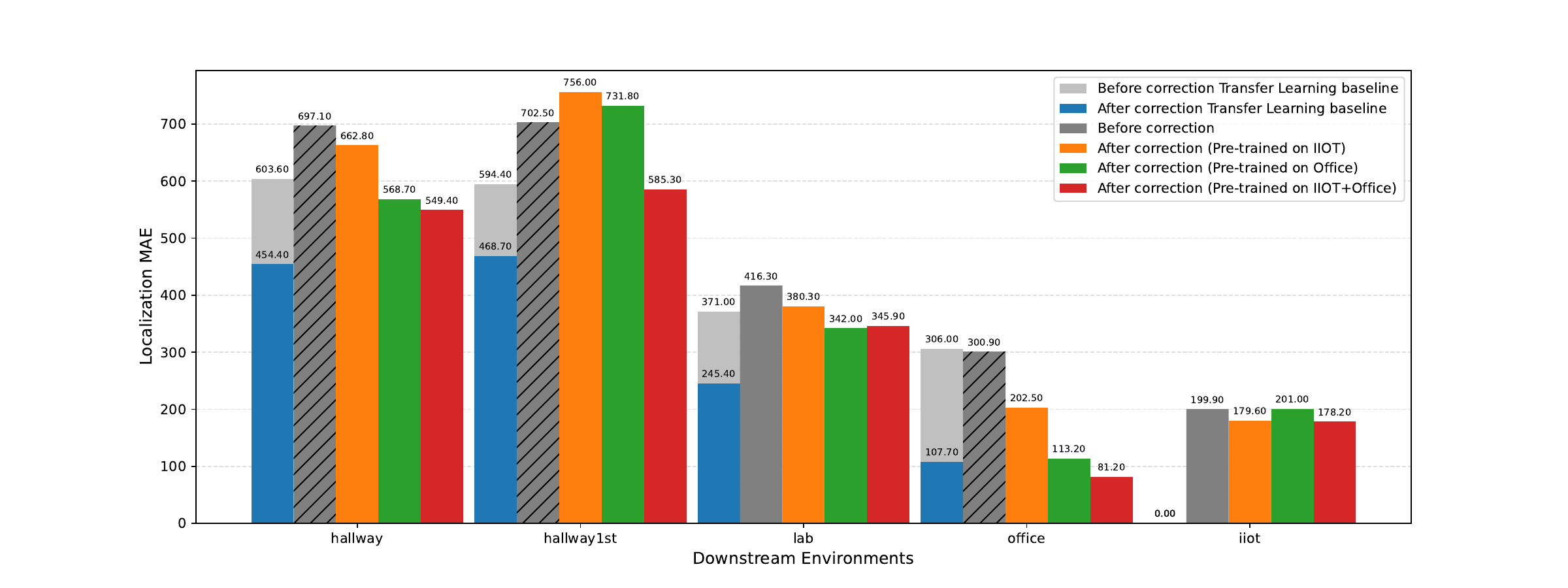}
        \caption{\Rev{Localization Mean Absolute Error (MAE) for the ranging error correction task.}}
        \label{fig:localization_env}
    \end{subfigure}

    \caption{\Rev{Performance of the fine-tuned foundation model on UWB downstream tasks, compared against a Transfer Learning baseline. The results show (a) classification accuracy and (b) regression MAE across five environments. Our model is evaluated with different pre-training datasets (IIOT, Office, or both). To test generalization, the Hallway, Hallway 1st Floor, and Lab environments were unseen during any pre-training phase.}}
    \label{fig:uwb_combined_results}
\end{figure*}

In figure~\ref{fig:lte_unseen}, the supervised setting achieves competitive performance, especially when the number of available samples is small. However, the self-supervised version outperforms all baselines as the number of samples increases, eventually surpassing 75\% accuracy with more than 300 samples. In comparison, Autoencoder + CNN reaches a plateau around 67\% accuracy, and the supervised iTransformer converges to slightly below 70\%.

In figure~\ref{fig:zigbee_unseen}, the Zigbee-excluded scenario, shows a more pronounced advantage for supervised fine-tuning. The supervised pre-training consistently achieves over 85\% accuracy with sufficient data, outperforming all baselines. The self-supervised variant starts at lower accuracy but gradually improves, eventually closing the gap with Autoencoder + CNN. While Autoencoder + CNN maintains steady but lower performance around 74\%, the iTransformer again remains below 72\%.

\Rev{Figure~\ref{fig:tsne} provides a t-SNE visualization of the feature embeddings learned by the self-supervised model after being fine-tuned for three distinct WTR downstream classification tasks. These plots illustrate the model's capability to learn discriminative representations for various wireless recognition scenarios. Subfigure~\ref{fig:a} presents the feature distribution for the short-range classification task, where the model produces well-defined clusters for LTE, Wi-Fi, DVB-T, and 5G, indicating effective separation despite potential spectral overlap. In Subfigure~\ref{fig:b}, which corresponds to the long-range classification task, the embeddings form distinct and compact clusters for 802.11ah, Zigbee, Lora, and Sigfox. This clear separation signifies superior classification performance, likely attributable to the unique signal characteristics of these sub-GHz technologies. Finally, Subfigure~\ref{fig:c} depicts the most challenging 8-class scenario, which combines all technologies. Even in this complex setting, the model successfully generates tangible clusters for the majority of the classes, demonstrating its robustness and its ability to learn effective features across a diverse set of wireless technologies.}

These results confirm that both supervised and self-supervised variants of our proposed model exhibit strong generalization to unseen technologies when fine-tuned with limited data. Supervised fine-tuning offers the highest performance in data-rich settings, whereas self-supervised initialization provides a notable advantage in low-data regimes. Autoencoder + CNN, while competitive, consistently underperforms as more samples become available. These findings underline the effectiveness of transformer-based models, particularly when pre-trained using self-supervised objectives, for recognizing unseen wireless technologies.

\Rev{
\subsection{UWB Fine-Tuning Results}
In this section, we evaluate the performance of our foundation model when fine-tuned for UWB downstream tasks. We assess its capabilities in two distinct scenarios: (N)LOS classification, a classification task, and ranging error correction, a regression task. The model's performance is compared with a specialized, state-of-the-art transfer learning study designed specifically for these UWB tasks in separate models. To test generalization, the evaluation includes environments that were entirely unseen during the pre-training phase, providing a clear measure of the model's adaptability.

Figure~\ref{fig:uwb_combined_results} presents the (N)LOS classification accuracy and localization MAE results, respectively. Our model was pre-trained using three different data configurations: the IIoT dataset alone, the Office dataset alone, or both, and then fine-tuned for the downstream environments. These results are compared against the Transfer Learning Baseline \cite{10195942}, which uses raw \ac{cir} data. A key distinction is that the baseline model is pre-trained in a supervised manner for a single task before knowledge is transferred. In contrast, our approach is based on a self-supervised foundation model designed to learn general representations applicable to multiple tasks, including WTR and UWB localization. Another difference lies in the size of the input data. The baseline model inputs 150 time samples of raw CIR. Our model, however, takes an input of 4096 time samples, where the 150 CIR samples are included and the other parts are zero-padded. This forces the model to identify and learn from sparse, localized features within a much larger input sequence.

In the unseen environments (Hallway, Hallway 1st Floor, and Lab), which were not included in the pre-training, our model demonstrates comparative performance as depicted in Figure~\ref{fig:nlos_env}, showing that our foundation model can be a valid alternative for task-specific AI solutions. For instance, when pre-trained on the Office dataset, our model achieves 81.00\% accuracy in the Hallway, surpassing the baseline's 76.00\%. Similarly, pre-training on the IIoT dataset yields 86.20\% accuracy in the Hallway 1st Floor, outperforming the baseline's 83.30\%. This highlights that the features learned during our model's self-supervised pre-training are highly transferable to new, unseen physical spaces.
For the seen environments (Office and IIoT), our fine-tuned model achieves high accuracy, reaching 100.00\% on the Office dataset (when pre-trained on Office) and 98.40\% on the IIoT dataset (when pre-trained on IIoT). This significantly exceeds the baseline's 92.80\% accuracy on the Office environment. Notably, the baseline shows 
0.00\% accuracy for the IIoT environment, as this dataset was used as its source domain for pre-training and was not evaluated as a separate downstream task in the original work.

\begin{figure}[h]
\centering
    \begin{subfigure}{0.49\textwidth}
        \includegraphics[width=\textwidth]{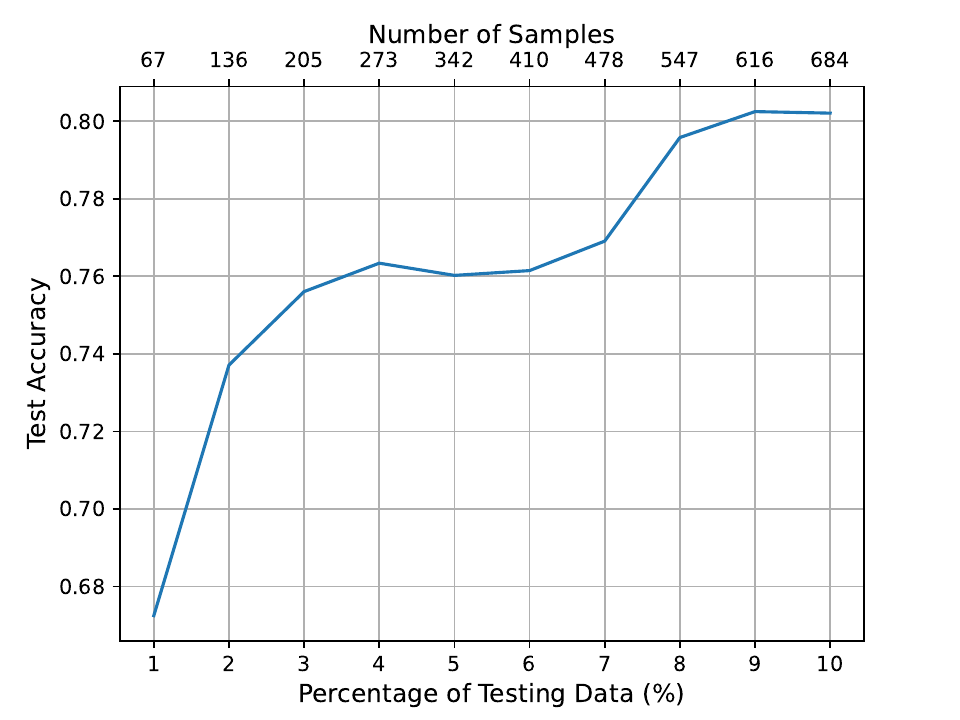}
        \caption{\Rev{(N)LOS classification accuracy}}
        \label{fig:sample_LOS}
    \end{subfigure}

    \begin{subfigure}{0.49\textwidth}
        \includegraphics[width=\textwidth]{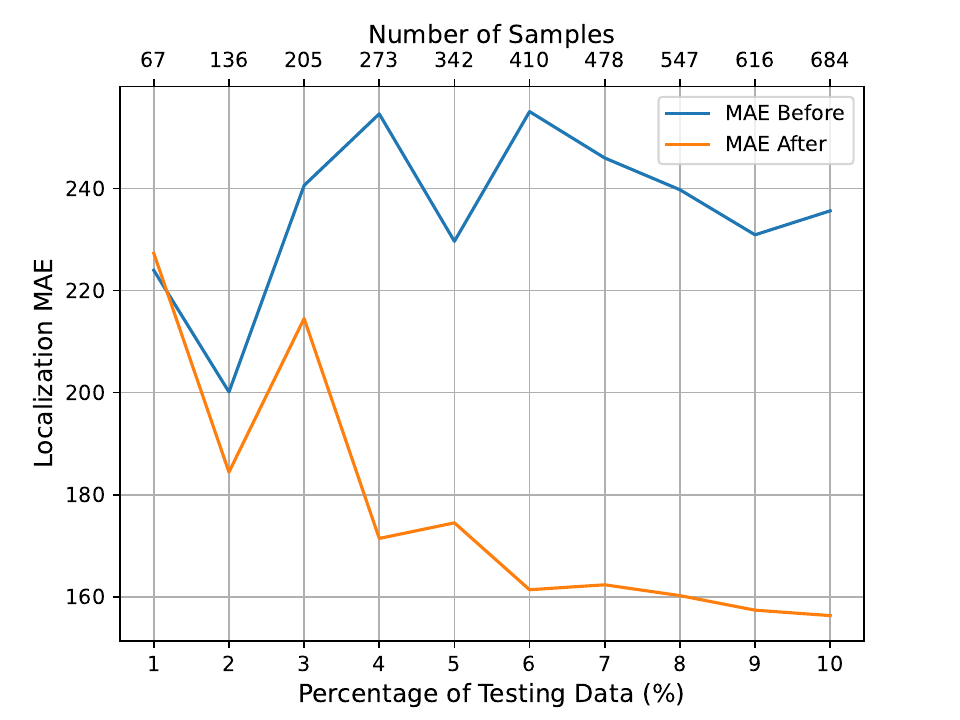}
        \caption{\Rev{Localization MAE before and after error correction}}
        \label{fig:sample_localization}
    \end{subfigure}
\caption{\Rev{Impact of the number of fine-tuning samples on UWB downstream task performance. These results are the average of the results for all environments, considering that all of them were seen in pre-training.}}
\label{fig:sample_analysis}
\end{figure}

For the UWB ranging error correction task, Figure \ref{fig:localization_env} compares localization MAE before and after correction. Hatched gray bars indicate the initial MAE, solid gray bars the MAE before transfer learning, and blue bars the transfer learning baseline’s results. Orange, green, and red bars show our fine-tuned foundation model using different pre-training datasets. Differences in reported errors (our model and baseline transfer learning) arise from random sample selection.

In unseen environments (hallway, hallway1st, lab), the non-foundation model baseline still often achieves lower MAE. For instance, in hallway1st it reduces MAE to 468.70 mm, while our best model achieves 585.30 mm. This is expected, as the baseline is optimized with a Bayesian framework for this specific task, whereas our model is a general-purpose foundation model adapted via self-supervised pre-training. Still, our model reduces error in all cases, showing the value of general features learned during pre-training. In seen environments (office, IIOT), our model is competitive or superior. In the office, pre-training on IIOT + Office yields 81.20 mm MAE, outperforming the baseline’s 107.70 mm. In IIoT, our model reduces the error from 199.90 mm to 178.20 mm, even though the baseline was not evaluated here.
These results highlight that fine-tuning a broad, self-supervised model can leverage rich general representations to achieve competitive performance.

To further investigate the data efficiency of our framework, we analyze the model's performance on both UWB tasks while varying the number of samples used for fine-tuning. Figure~\ref{fig:sample_analysis} depicts the performance with data availability, from 1\% (67 samples) to 10\% (684 samples) of the total data from all environments. For the (N)LOS classification task in Figure~\ref{fig:sample_LOS}, the test accuracy shows a clear positive correlation with the number of fine-tuning samples. The model's accuracy climbs from just under 70\% with only 67 samples to over 80\% when fine-tuned with 684 samples. This robust improvement highlights the model's ability to effectively use additional data to enhance its classification accuracy when the data includes information from different environments.

A similar trend is observed for the error correction task. Figure~\ref{fig:sample_localization} shows that the corrected error is consistently lower than the initial error, demonstrating the model's benefit regardless of sample size. As the number of fine-tuning samples increases, the localization MAE decreases. With just 2\% of the data (136 samples), the model already achieves a significant error reduction from over 200 mm to nearly 180 mm. This demonstrates that even a small, targeted dataset is sufficient to effectively adapt the foundation model for the regression task.

}
\subsection{Complexity Analysis}

The original Transformer architecture \cite{NIPS2017_3f5ee243} exhibits a computational complexity of $O(N^2)$ in both time and space, where $N$ represents the number of input tokens. In timeseries modeling tasks like ours, $N$ equals the input sequence length $L$. This quadratic complexity poses a significant bottleneck when dealing with long sequences, as it leads to substantial computational time and memory consumption.

To address this challenge, we used a patching technique that reduces the complexity of the model. By segmenting the input sequence into patches of length $P$ with a stride $S$, the effective number of tokens becomes $N \approx L/S$. This method changes computational complexity to $O\left((L/S)^2\right)$, reducing the quadratic growth.

\begin{table}[h]
\centering
\renewcommand{\arraystretch}{1.4}
\begin{tabular}{c|c|c|c}
\textbf{Patch} & \textbf{Accuracy (\%)} & \textbf{Time (s/epoch)} & \textbf{Memory (MB)} \\
\hline \hline
8     & 99.42 & 201.2  & 8960 \\
\hline
32    & 99.62 & 47.54  & 1900 \\
\hline
64    & 99.79 & 45.15  & 1210 \\
\hline
\textbf{128}   & \textbf{99.55} & \textbf{44.14}  & \textbf{980} \\
\hline
256   & 98.91 & 44.60  & 850 \\
\hline
512   & 86.99 & 44.42  & 800 \\
\hline
1024  & 75.33 & 44.30  & 770 \\
\end{tabular}
\caption{Effect of patch size on classification accuracy, training time per epoch, and GPU memory usage.}
\label{tab:patch_size_analysis}
\end{table}

\begin{figure}[h]
\centering
\includegraphics[width=0.53\textwidth]{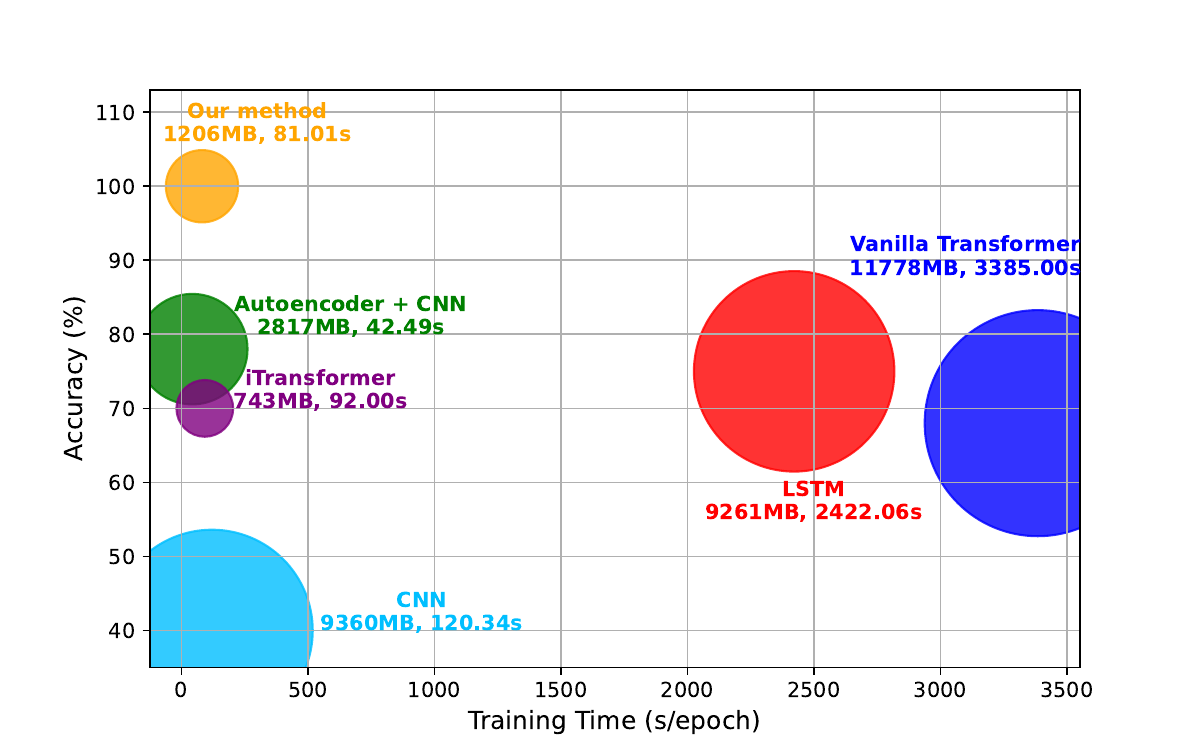}
\caption{
        \Rev{Accuracy vs. training time per epoch for different models, with bubble size representing memory usage. Our method achieves the highest accuracy with low training time and memory footprint, outperforming baseline models in all three aspects.}
    }
\label{fig:complexity_comparison}
\end{figure}

To further analyze the impact of patching on efficiency, we evaluate the effect of varying patch sizes $P$ on classification accuracy (assuming that the stride is equal to the patch size in this evaluation), training time per epoch, and GPU memory usage, as shown in Table~\ref{tab:patch_size_analysis}. As the patch size increases, we observe a drop in computational cost. For example, increasing $P$ from 8 to 128 reduces the training time by over 75\% (from 201.2s to 44.14s) and memory usage by almost 9$\times$ (from 8960MB to 980MB), while maintaining a high classification accuracy above 99.5\%. 

However, very large patch sizes (e.g., $P=512$ or $P=1024$) lead to a significant degradation in accuracy, indicating a loss of important temporal resolution. To balance this trade-off between efficiency and performance, we selected a patch size of $P=128$ as the optimal configuration. It achieves 99.55\% accuracy while minimizing both training time and GPU memory usage.

Figure~\ref{fig:complexity_comparison} presents the complexity comparison results for the 8-class pre-training scenario, corresponding to the accuracies reported in Table~\ref{8class:pretrain}. Our method achieves the highest accuracy while maintaining significantly lower training time (81.01 seconds per epoch) and minimal GPU memory usage (1206MB), clearly outperforming all baseline models. In contrast, models such as the Vanilla Transformer and LSTM exhibit extremely high computational costs, with training times of 3385s and 2422s per epoch and memory usage of 11778MB and \Rev{9261MB}, respectively—yet they achieve much lower accuracies of 67.6\% and \Rev{75.62}\%.

Even efficient models such as Autoencoder + CNN or iTransformer deliver lower accuracy. This clear performance gap highlights the efficiency and scalability of our patch-based Transformer. By reducing the effective sequence length through patching, our model avoids the $O(L^2)$ time and memory complexity typical of full-length Transformer attention, achieving a more practical $O((L/S)^2)$ complexity with minimal compromise in performance.

\section{Conclusion \& Future Works} \label{sec:conclusion}

This work introduced a patching-based Transformer approach for \ac{wtr}, aiming to overcome the limitations of baseline models in generalization to unseen wireless technologies and adaptability across different sampling rates. The proposed foundation model achieved higher classification accuracy than traditional methods while maintaining a significantly lower computational complexity. Unlike standard Transformer models, which suffer from high training times, our approach offers efficiency comparable to lightweight CNN-based methods. Evaluations were done on a heterogeneous dataset containing eight wireless technology classes with varying sampling rates. With each supervised and self-supervised pre-training combined with fine-tuning, the foundation model consistently achieved strong generalization performance, demonstrating its suitability for real-world applications.

Future work will focus on integrating more diverse and real-world datasets to further assess the model’s scalability and robustness. Deploying the model in live communication systems will help bridge the gap between experimental validation and practical deployment, ensuring stable operation in dynamic wireless environments. Additionally, inspired by the concept of foundation models \cite{10615509}, we aim to extend this framework into a unified wireless physical-layer model capable of handling multiple tasks such as interference detection and modulation classification. As a complementary direction, we also plan to explore the use of agentic AI systems to autonomously adapt and optimize wireless technology recognition in evolving environments, further enhancing the model’s applicability and resilience.

\bibliographystyle{unsrt}
\bibliography{references}
\end{document}